\documentclass[twocolumn, secnumarabic, amssymb, nobibnotes, aps, prd, floats, floatfix]{revtex4-1}

\usepackage{graphicx}
\usepackage[export]{adjustbox}
\usepackage{amsmath}
\usepackage{color}

\begin{document}
\title{A stochastic root finding approach: The Homotopy Analysis Method applied to Dyson-Schwinger Equations}%

\author{Tobias Pfeffer}
\author {Lode Pollet}
\affiliation{Department of Physics, Arnold Sommerfeld Center for Theoretical Physics,
University of Munich, Theresienstrasse 37, 80333 Munich, Germany}
\date{\today}%

\begin{abstract}
We present the construction and stochastic summation of rooted-tree diagrams, based on the expansion of a root finding algorithm applied to the  Dyson-Schwinger equations (DSEs). The mathematical formulation shows superior convergence properties compared to the bold diagrammatic Monte Carlo approach and the developed algorithm allows one to tackle generic high-dimensional integral equations, to avoid the curse of dealing explicitly with high-dimensional objects and to access non-perturbative regimes. The sign problem remains the limiting factor, but it is not found to be worse than in other approaches. We illustrate the method for $\phi^4$ theory but note that it applies in principle to any model.
\end{abstract}

\maketitle

\section{Introduction}
%
%
Developing first-principles methods for strongly-interacting many-body systems remains an active field of research in theoretical physics. Quantum Monte Carlo algorithms  are often the method of choice because of their versatility: Unprecedented insight has been gained with path integral Monte Carlo methods for (bosonic) cold atomic systems \cite{RecentDevelopmentsPIMC, ComparisonMCExperiment, HiggsBoseHubbard}, superfluid and supersolid $^4$He \cite{FateVacancySuperSolid, SFGrainBoundaries}. Determinant Monte Carlo simulations remain indispensable for the nuclear shell model \cite{NuclearShell}, lattice quantum chromodynamics \cite{HybridMonteCarlo}, fermions at unitarity \cite{UnitaryFermionsI, UnitaryFermionsII}, the Hubbard model \cite{detQMCLatticeFermions}, topological insulators \cite{KaneMeleHubbardAssad, KaneMeleHubbardWu}, and currently certain designer Hamiltonians with gauge fields \cite{GaugeFields} have been added to this list. They are also used as impurity solvers in dynamical mean-field theory \cite{ImpuritySolvers}. Diffusion Monte Carlo \cite{DiffusionMonteCarloElectronGroundState} and full configuration interaction quantum Monte Carlo \cite{FICQMCcode, FICQMCapplication} are often used for electronic ground state properties and chemical molecules.\\
%
%
However, in the absence of a positive expansion scheme Monte Carlo algorithms scale exponentially. This is the case for interacting Fermi systems without special symmetry and frustrated spin-systems. Frustration can lead to bad sampling properties already for classical models. It is generally accepted that this exponential scaling is unlikely to be overcome in general \cite{SignProblem}. One hence tries to develop methods such that the physics can be retrieved before the exponential scaling makes the calculations impossible.\\
%
%
Diagrammatic Monte Carlo (diagMC) is based on the stochastic evaluation of the perturbative Feynman series and recently gained attention: Its greatest appeal is that, for generic fermionic problems, the inevitable sign problem does not lead to an exponential scaling in the system volume but in the expansion order. This typically enables the evaluation of the diagrams up to order 5-10 depending on the problem under consideration, from which the converged answer can hopefully be extracted. 
When taking into account the range of problems, starting with the polaron problem \cite{FermiPolaron, FermiPolaron3DMass, FermiPolaron2DVlietinck} over models of frustrated spins \cite{FrustratedSpins,SpinIce} and to strongly correlated electron systems \cite{Hubbard, AnisotropicHubbard, PseudoGapHubbard, ColoumbElectronPhonon, DiracLiquid}, this shows that diagMC is a flexible and quite universal tool. Recently, it has also been used to systematically reintroduce non-local correlations in the dynamical mean-field theory framework \cite{DualFermionGull, DualFermionGukelberger}.\\
%
%
One of the biggest challenges for diagMC is the issue of a possible zero convergence radius for which a necessary condition is Dyson's collapse argument: The Feynman series corresponds to a Taylor series in the coupling constant, and viewing the latter as a complex variable,  the system is seen to be unstable against collapse with a zero convergence radius as a result. Dyson formulated the argument originally for quantum electrodynamics, but it applies generally to any bosonic system, in particular to $\phi^4$ theory which is studied in this paper. 
%
%
An alternative approach is provided by the skeleton technique known as Hedin's equations in material science or the Dyson-Schwinger equations (DSEs) in physics.
If one can prevent the perturbative expansion of the DSEs (which, in case of $\phi^4$ theory, would bring us back to Feynman diagrams subject to Dyson's collapse~\cite{Phi4Buividovich}) and instead solve these high-dimensional integral equations directly, progress is possible.\\
%
%
The main result of this paper is to present an algorithm which can be used to solve such high-dimensional integral equations stochastically. The main technique we develop in this paper is the expansion and stochastic summation of the Homotopy Analysis Method (HAM)~\cite{HAMbook}. The HAM is a numerical method to find solutions of non-linear differential and integral equations with a growing number of applications in science, finance, fluid dynamics and engineering~\cite{HAMbook2, HAMbook3}. Therefore, series convergence problems, such as Dyson's collapse, are avoided if this algorithm is applied to the DSEs as long as the HAM is able to find the solutions of the DSEs. A crucial aspect is that we can prevent the explicit storage and manipulation of $n-$leg vertices (as long as only integrals over such quantities are needed), which for $n \ge 4$ is all but impossible currently. This overcomes the main limitation of numerical methods dealing with DSEs or related skeleton techniques, e.g., functional renormalization group approaches \cite{fRGStatMech, fRGFermi}. 
We note in passing that the recently developed technique of Grassmannization \cite{GrassmanizationIsing} also prevents Dyson's collapse.
\\
\\
The paper is organized as follows, featuring an increasing degree of complexity ranging from a zero-dimensional problem to quantum field theory: In the next chapter (Sec.~\ref{sec:two}) we illustrate the key ideas for a coupled set of algebraic equations and solve these issues by introducing the zero-dimensional equivalent of rooted trees. We proceed with discussing the Homotopy Analysis Method mathematically and illustrate it for the case of a one-dimensional integral equation in Sec.~\ref{sec:three}. Next, we apply the technique to the $\mathcal{Z}_2$ symmetric $\phi^4$ model in 1D (Sec.~\ref{sec:four}). Sec.~\ref{sec:three} and \ref{sec:four} work out the technical implementations of the ideas presented in Sec.~\ref{sec:two} and can therefore be skipped by readers who are not interested in the detailed implementation. As a non-trivial example we apply the method to the $\mathcal{Z}_2$ symmetric $\phi^4$ model in 2D (Sec.~\ref{sec:five}). We conclude and provide an outlook in Section~\ref{sec:conclusion}. Appendix A contains further details of the toy model of Sec.~\ref{sec:two} whereas Appendix B shows how the ideas developed in Sec.~\ref{sec:two} can be formulated for the full DSEs, including the integro-differential equation for the vertex function, showing that the developed algorithm can, in principle, be applied to any model.
%
%
%
\section{Solution Strategy }
\label{sec:two}
%
The key idea can already be understood from the $0$ space-time dimensional case of the $\phi^4$ model with action
\begin{equation}
S_E(\phi) = \frac{1}{2}\phi^2+\frac{\lambda}{4!}\phi^4 \, .
\end{equation}
In this case the field is reduced to just a single variable $\phi$ and the connected n-point function $G^{(n)}_c(\mathbf{x}_1,\dots,\mathbf{x}_n)$ reduces to the cumulant $\kappa_n$ which is obtained from $\kappa_n = \left.\frac{ \text{d}^n F(J)}{ \text{d} J^n}\right|_{J=0}$ where
\begin{equation}
F(J)=\log \langle e^{J\phi} \rangle = \log \int \text{d}\phi \, e^{-S_E(\phi)+J\phi} . 
\end{equation}
The differential form of the DSEs can be written as
\begin{equation}
\frac{\text{d} S_E}{\text{d} \phi}\left[\frac{ \text{d} }{\text{d} J}+\frac{\text{d} F(J)}{\text{d} J}\right]=J\, .
\end{equation}
From this form of the DSEs the first two non-zero cumulants can be derived by differentiating with respect to $J$,
\begin{equation}
\begin{split}
\label{truncatedset0d}
\kappa_2+\frac{\lambda}{6}\kappa_4+\frac{\lambda}{2}\kappa_2^2&=1 \\
\kappa_4+\frac{\lambda}{6}\kappa_6+2\lambda\kappa_2\kappa_4+\lambda\kappa_2^3&=0.
\end{split}
\end{equation}
This builds an infinite tower of non-linear equations with an infinite number of unknown variables ($\kappa_2$ depends on $\kappa_2$ and $\kappa_4$; $\kappa_4$ depends on $\kappa_2, \kappa_4$ and $\kappa_6$, etc.), known as the (integral) DSEs.\\
%
%
The infinite tower can be perturbatively expanded in the coupling constant $\lambda$ (cf. Appendix \ref{kappa2lambdaexp}),
\begin{equation}
\begin{split}
\kappa_2 =& \, 1 - \frac{\lambda}{2}\kappa_2^2 - \frac{\lambda}{6}F^{\text{pert}}(\kappa_2) \\
F^{\text{pert}}(\kappa_2) =& - \lambda \kappa_2^3 + 2 \lambda^2  \kappa_2^4 -\frac{17}{3} \lambda^3 \kappa_2^5 + \mathcal{O}(\lambda^4).
\end{split}
\label{towerlambdaexpanded}
\end{equation}
Such an expansion is in close analogy with existing bold diagrammatic Monte Carlo codes which presently rely on the Luttinger-Ward functional \cite{LuttingerWard} (or a closely related functional). In these approaches the Luttinger-Ward functional is constructed as the sum of all possible closed diagrams built of (full) 2-point correlation functions and which do not fall apart when cutting any two 2-point correlation function lines. We shall refer to such an expansion as a skeleton series. These are, however, often asymptotic. \\
%
%
In contrast to such a perturbative expansion another strategy is to truncate the infinite tower, e.g. by setting $\kappa_6=0$ (this also assumes that 4-point vertices (cf. $\kappa_4$) can be dealt with appropriately). The system\,(\ref{truncatedset0d}) then yields a closed set of two equations.
\begin{figure}
\centering
\includegraphics[width=0.45\textwidth]{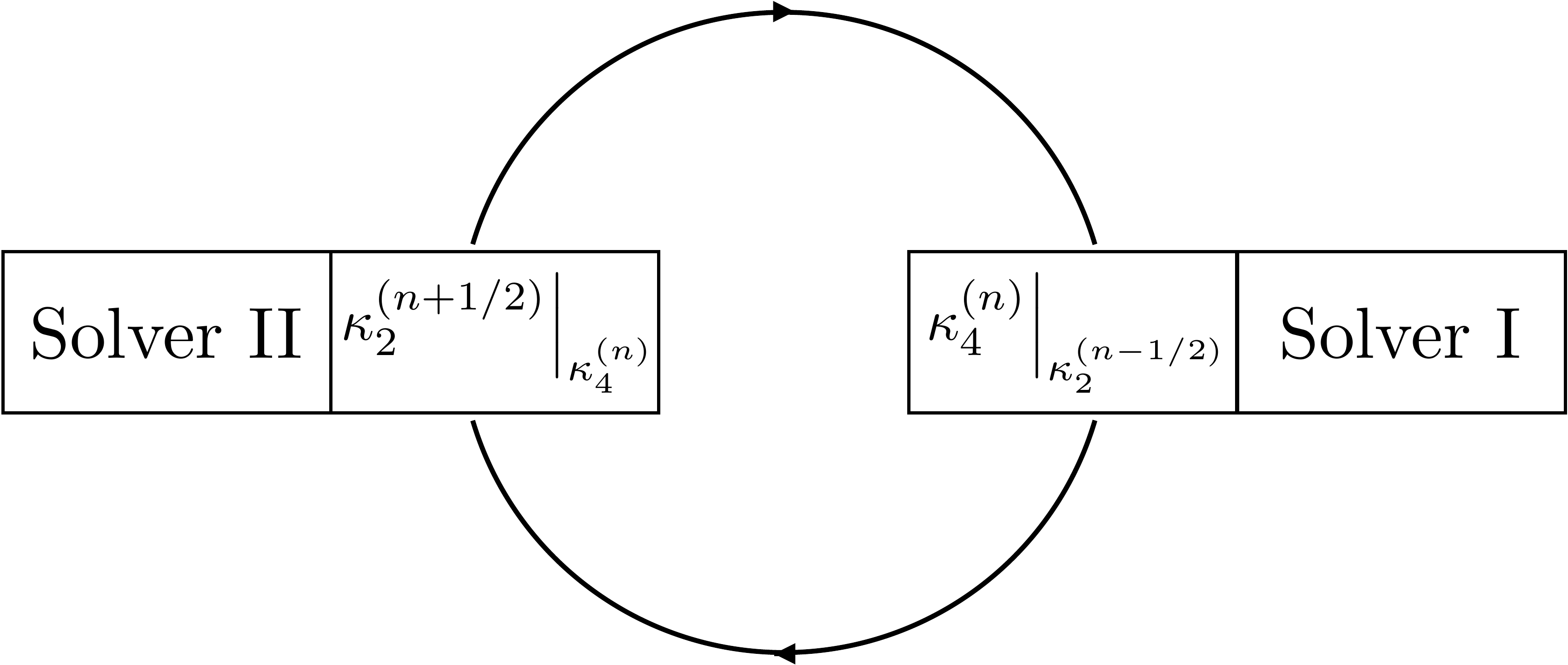}
\caption{The self-consistency loop to solve the coupled set of DSEs. The solution of the coupled equations, cf. Eqs.~(\ref{solverI}) and~(\ref{solverII}), after the $n$-th iteration in the self-consistency loop is denoted as $\kappa^{(n)}_4$, $\kappa^{(n+1/2)}_2$.}
\label{fig:selfconsistencyloop}
\end{figure}
The usual procedure to solve such a system stochastically is by fixed point iterations: Starting from initial guesses $\kappa_2^{(0)}$ and $\kappa_4^{(0)}$ one obtains in iteration $n+1$ an approximate solution which depends on the solution in step $n$:
\begin{eqnarray}
\kappa_2^{(n+1)} & = & 1 - \frac{\lambda}{6}\kappa^{(n)}_4 - \frac{\lambda}{2} (\kappa^{(n)}_2)^2 \nonumber \\
\kappa_4^{(n+1)} & = &  - 2 \lambda \kappa^{(n)}_2\kappa^{(n)}_4 - \lambda (\kappa^{(n)}_2)^3.
\label{}
\end{eqnarray}
Generically, such a fixed point iteration may not have a stable fixed point. It can be checked that already for $\lambda \sim 2$ the above fixed point iteration is diverging.
To improve stability we consider each of the equations as non-linear (implicit) equations which are connected by the self-consistency loop shown in Fig.\;\ref{fig:selfconsistencyloop},
\begin{eqnarray}
\kappa^{(n)}_4 & = & - 2\lambda\kappa^{(n-1/2)}_2\kappa^{(n)}_4 - \lambda\left(\kappa_2^{(n-1/2)}\right)^3   \label{solverI}  \\
\kappa^{(n+1/2)}_2 & = & 1 - \frac{\lambda}{6}\kappa^{(n)}_4- \frac{\lambda}{2}\left(\kappa^{(n+1/2)}_2\right)^2. \label{solverII}  
\end{eqnarray}
The equations (\ref{solverI}), (\ref{solverII}) are solved by considering each as a root finding problem  $f  (\kappa^{(n)}_4) =  0$, $g (\kappa^{(n+1/2)}_2) =  0$. In Fig.\;\ref{fig:selfconsistencyloop} ``Solver I" denotes a root finding algorithm which solves $f (\kappa^{(n)}_4) =  0$ for $\kappa^{(n)}_4$ with fixed $\kappa^{(n-1/2)}_2$ while ``Solver II" solves $  g (\kappa^{(n+1/2)}_2) =  0$ for $\kappa^{(n+1/2)}_2$ with fixed $\kappa^{(n)}_4$.
\\
In order to find the root of  $f (\kappa^{(n)}_4) =  0$ ``Solver I"  can  employ the Newton-Raphson method which will be substituted by the Homotopy Analysis Method (HAM) in cases where the DSEs have the form of integral equations (cf. Sec.~\ref{sec:three}). The Newton-Raphson method finds the root corresponding to the procedure,
\begin{equation}
\kappa^{(n)}_{4,i+1} = \kappa^{(n)}_{4,i}+\frac{f ( \kappa^{(n)}_{4,i} ) }{ f ^ \prime ( \kappa^{(n)}_{4,i} ) }. 
\label{newtonraphson}
\end{equation} 
The index $i$ denotes the iteration number in the auxiliary Newton-Raphson process and is distinct from the index $n$ for the iterations in the main self-consistency problem (i.e., the DSEs).
The quality of the solution generically improves with the number of iterations $i$ but the result of each iteration step has to be stored in order to use it as a starting point for the next iteration step.\\
In case of a field theory, $\kappa_4$ should be thought of a hard-to-manipulate connected 4-point correlation function $G_c^{(4)}(\mathbf{x_1},\mathbf{x_2},\mathbf{x_3},\mathbf{x_4})$.
To overcome the storage problem, we use a recursive expression for the $(i+1)$-th approximation to the root,
\begin{equation}
\begin{split}
\kappa^{(n)}_{4,i+1} &= \kappa^{(n)}_{4,i}+\frac{  f  ( \kappa^{(n)}_{4,i} ) }{ f ^ \prime ( \kappa^{(n)}_{4,i} ) }
 \\& = \kappa^{(n)}_{4,i-1} + \frac{  f ( \kappa^{(n)}_{4,i-1} ) }{ f^\prime( \kappa^{(n)}_{4,i-1} ) }  + \\
 & \hspace{9mm} + \frac{  f ( \kappa^{(n)}_{4,i-1} + \frac{  f ( \kappa^{(n)}_{4,i-1} ) }{ f^\prime( \kappa^{(n)}_{4,i-1} ) } ) }{ f ^\prime( \kappa^{(n)}_{4,i-1} + \frac{  f ( \kappa^{(n)}_{4,i-1} ) }{ f ^\prime( \kappa^{(n)}_{4,i-1} ) } ) }\\
 &= \dots = F^{\text{tree}} ( \kappa^{(n)}_{4,0} ).
 \label{reconstructNewton}
\end{split}
\end{equation}
The expanded result for $\kappa^{(n)}_{4,i+1}$ is a real-valued function $F^{\text{tree}} ( \kappa^{(n)}_{4,0} )$ depending on the initial guess $\kappa^{(n)}_{4,0}$ for the root finding. It also depends implicitly on $\kappa^{(n-1/2)}_2$ as this cumulant was taken fixed in the root finding of ``Solver I" and can therefore be viewed as another parameter. In the following such an expansion is referred to as a tree expansion.
\\
We can now proceed with ``Solver II" using the tree expansion $F^{\text{tree}}$, which yields
\begin{equation}
\kappa^{(n+1/2)}_2+\frac{\lambda}{6} F^{\text{tree}} (\kappa^{(n)}_{4,0}) + \frac{\lambda}{2} (\kappa^{(n+1/2)}_2) ^2=1.\, 
\label{combinedsolverexpansion}
\end{equation}
``Solver II" is used to solve for the unknown $\kappa_2^{(n+1/2)}$ after which iteration step $n$ is finished. Since the prime object of ``Solver II" (i.e., the 2-point correlation function) is easy to store and manipulate, ``Solver II" does not require a tree expansion.
This notation makes clear that we can combine both solvers, shown as ``Solver I+II" in Fig.\,\ref{fig:combinedsolver}, whenever  the root finding of ``Solver I" has been expanded in terms of the function $F^{\text{tree}}$. Compared to Fig.\,\ref{fig:selfconsistencyloop} the only change is that $\kappa_4$ is represented by a tree expansion and not by a single object.\\
%
%
By dropping the self-consistency index $n$ the final result of (\ref{combinedsolverexpansion}) can be compared with the skeleton series expansion of (\ref{towerlambdaexpanded}). The tree expansion inherits the properties of ``Solver I" as it is constructed to exactly represent the root finding algorithm. Therefore, by using the tree expansion the often asymptotic skeleton series expansion of bold diagrammatic Monte Carlo is avoided and $F^{\text{tree}}$ is used instead of $F^{\text{pert}}$. It should also be noted that once the infinite tower is truncated by setting $\kappa_n = 0$ for some fixed $n$ the perturbative expansion of the truncated tower (cf. Eq. (\ref{towerlambdaexpanded})) has a finite convergence radius, e.g., for $\kappa_6=0$ the convergence radius $r$ is given by $|r|< 2 \lambda \kappa_2$ (cf. (\ref{solutionexpand})).
\begin{figure}[t]
\centering
\includegraphics[width=0.45\textwidth]{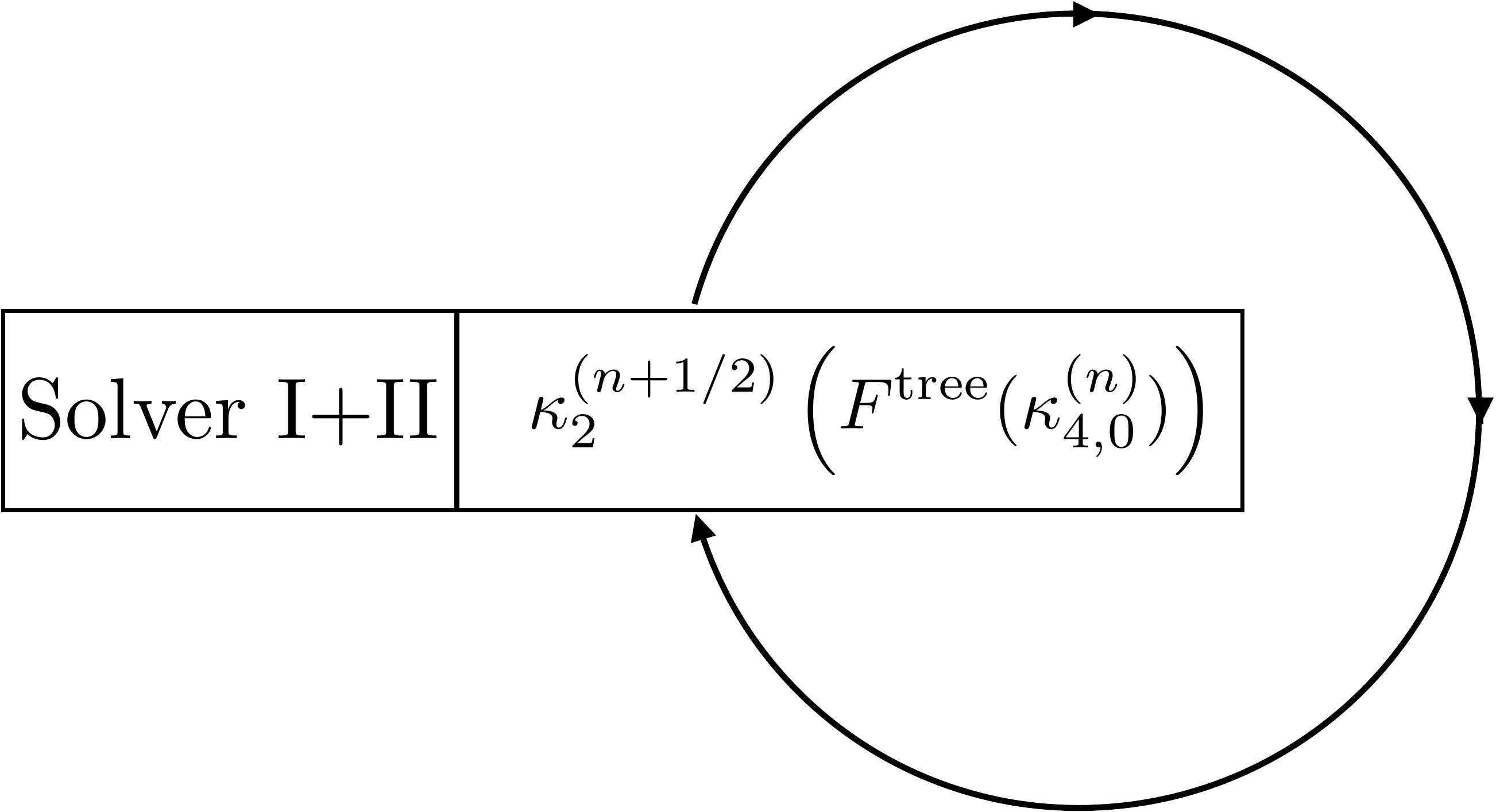}
\caption{``Solver I" and ``Solver II" from Fig~\ref{fig:selfconsistencyloop} can be combined into ``Solver I+I" here. ``Solver I+II" is in fact ``Solver II" where $\kappa_4$ is represented by the tree expansion of  ``Solver I", cf. Eq.~(\ref{combinedsolverexpansion}).}
\label{fig:combinedsolver}
\end{figure}
\\
%
%
We have glossed over the computation of the expansion $F^{\text{tree}}$ in Eq.~(\ref{reconstructNewton}), which requires keeping track and summing all the terms generated in the tree expansion to all orders in the recursion and which constitutes a formidable problem for a finite dimensional field theory.
Here, we propose a stochastic approach to the tree expansion, in the same spirit as diagrammatic Monte Carlo samples Feynman diagrams. Consequently, the configuration space is given by a collection of diagrams where every diagram is in one to one correspondence to a single term in the tree expansion and a weight that is the product of the individual building blocks. 
How this Monte Carlo average is exactly implemented in practice is subject of the next section.
%
%
\section{Integral equations and Diagrammatic Monte Carlo}
\label{sec:three}
The aim of this section is to arrive at a practical scheme for the generalization of the tree expansion (\ref{reconstructNewton}). The mathematical framework is provided by the Homotopy Analysis Method (Sec.~\ref{sec:HAM}). A Monte Carlo updating scheme is presented in Sec.~\ref{sec:HAM_updates}, and finally, the method is illustrated for a one-dimensional integral equation where the answer can be compared with the exact one, see Sec.~\ref{sec:HAM_example}.
\subsection{Homotopy Analysis Method}
\label{sec:HAM}
\begin{figure}
\centering
\includegraphics[width=0.4\textwidth]{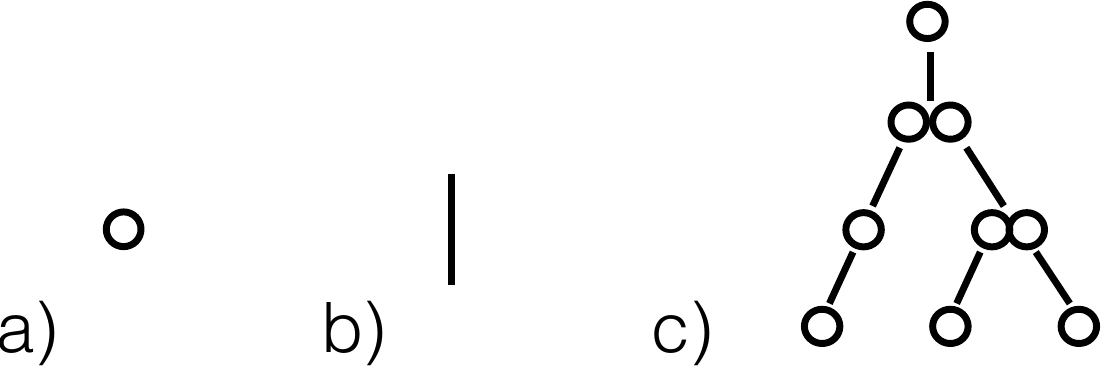}
\caption{The basic elements of the rooted trees. a) The roots and leafs are drawn as open circles. b) Branches are connecting the roots with the leafs. c) A set consisting of a root, leafs and branches constitutes a rooted tree.}
\label{fig:rootedtreeelements}
\end{figure}

\begin{figure}
\centering
	\includegraphics[width=0.4\textwidth]{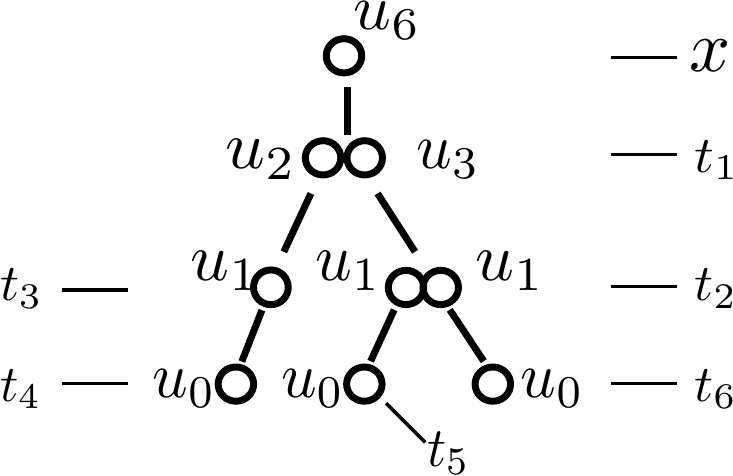}
\caption{Each fully grown rooted tree corresponds to an integral expression, e.g. Eq.\,(\ref{fullexpression}), in the tree expansion which can be read off from the labeled rooted tree.}
\label{fig:fullgrowntree}
\end{figure}

We are interested in finding the roots of a non-linear equation $\mathcal{N}$,
\begin{equation}
\mathcal{N}[f]=0.
\end{equation}
The idea of the Homotopy Analysis Method (HAM) \cite{HAMbook} is to rewrite this problem with the help of an embedding parameter $q$ and a convergence control parameter $h$,
\begin{equation}
 (1-q)\mathcal{L}[\phi(x,q)-f_{0}] + q h\mathcal{N}[\phi(x,q)]=0.
\label{eq:HAM_sys}
\end{equation}
Here, $\mathcal{L}$ is an arbitrary linear operator with $\mathcal{L}[0]=0$. This equation is called the $0$-th order deformation equation.
By setting $q=0$ and $q=1$ one sees that the initial guess for the solution of the root finding problem $f_0$ is transformed to the full solution $f(x)=\phi(x,q=1)$.
Under the assumption that this transformation is smooth and the Taylor expansion is well-defined, one can write
\begin{equation}
 \phi(x,q)=\sum_m \left.\frac{1}{m!}\frac{\text{d} ^{m}}{\text{dq}^{m}}\phi(x,q)\right|_{q=0}q^{m}.
\end{equation}
Therefore,
\begin{equation}
\begin{split}
 f(x)=\phi(x,q=1)= \sum_m u_{f,m} \\
\text{with} \hspace{3mm} u_{f,m} = \left.\frac{1}{m!}\frac{\text{d} ^{m}}{\text{dq}^{m}}\phi(x,q)\right|_{q=0}  . 
 \end{split}
\end{equation}
The Taylor coefficients $u_{f,m}$ can be obtained by differentiating the $0$-th order deformation equation $m$ times  with respect to $q$ and setting $q=0$ afterwards. This can be done analytically yielding a set of deformation equations. It can also be represented graphically with the help of diagrams as will be shown in the following. \\
The non-linear equation under consideration typically has the form
\begin{equation}
 f(x)=c(x)+\int^{b}_{a}K(x,t)n\left( f(t) \right) dt,
\end{equation}
with given  functions $c$, $n$ and the kernel of the integration $K$. Choosing the HAM convergence parameters $h=1$ and the linear operator $\mathcal{L}$ as the identity operator, the $0$-th order deformation equation is given by
\begin{equation}
 (1-q) \left[ \phi(x,q) - f_0(x) \right] + q \mathcal{N}[ \phi(x,q) ] = 0,
\end{equation}
where
\begin{equation}
 \mathcal{N}[ \phi(x,q) ] = \phi(x,q) - c(x) - \int^{b}_{a} K(x,t) n(\phi(t,q)) \text{d} t.
\end{equation}
Taking the derivative with respect to $q$ $m$ times in the $0$-th order deformation equation and taking $q=0$ afterwards gives an equation for $u_m=\frac{1}{m!}\frac{\text{d}^m \phi(x,q) }{\text{dq}^m}|_{q=0}$.
For $m\geq2$ one finds
\begin{equation}
u_{f,m}(x) = \frac{1}{(m-1)!}\left.\frac{ \text{d} ^{m-1}}{\text{dq}^{m-1}}\int^{b}_{a}K(x,t)n(\phi(t,q)) \text{d} t\right|_{q=0}.
\label{eq:u}
\end{equation}
This is the starting point for the tree expansion of the HAM. The above equation can be written without specifying the non-linear function $n$,
\begin{align}
 u_{f,m} & (x)  =   \frac{1}{(m-1)!}\int^{b}_{a} \text{d} t \; K(x,t) \left.\frac{\text{d} ^{m-1} n(\phi(t,q))}{\text{dq}^{m-1}} \right|_{q=0}  \nonumber \\
  = &  \frac{1}{(m-1)!}  \int^{b}_{a} \text{d} t \; K(x,t) \sum_{k=1}^{m-1} n^{(k)}(\phi(t,q))   \nonumber \\
	& \times \left.  B_{m-1,k}\left(\phi(t,q)^\prime,\dots,\phi(t,q)^{(m-k-1)}\right)\right|_{q=0}  \label{faa}  \\
 = &  \frac{1}{(m-1)!}  \int^{b}_{a} \text{d} t \; K(x,t)  \sum_{k=1}^{m-1} n^{(k)}(u_0(t)) \nonumber \\
	&  \times B_{m-1,k} \left( u_{f,1}(t),\dots,(m-k-1)! u_{f,m-k-1}(t) \right) . \nonumber
\end{align}
Here $B_{m-1,k}$ are the Bell polynomials of the second kind \cite{BellPolynomialRef} encoding the combinatorial coefficients produced by the action of the derivative $\frac{\text{d} ^{m-1}}{\text{d} q^{m-1}}$ on $n( \phi(t,q) )$. It can be checked that for the initial guess of the root finding $f_0(x)=u_{f,0}(x)=c(x)$ Eq.\,(\ref{faa}) also holds for $m=1$.\\
When the root finding algorithm has converged in step $M$, all references to the previous iteration results $u_{f,m}$, $m<M$ should be eliminated from Eq.\,(\ref{faa}) by using Eq.\,(\ref{faa}) recursively for every $u_{f,m}$, $m\neq0$, leading to the desired generalization of Eq.~(\ref{reconstructNewton}). This is again the general procedure leading to the tree expansion.
To avoid confusion it should be noted that even though the solution of the non-linear equation is found in the form of $\sum_m u_{f,m}$ by the HAM this expansion does not constitute the tree expansion as in order to calculate $u_{f,j}$ all $u_{f,m}$ with $m<j$ have to be calculated and stored, cf. (\ref{faa}).
In the sequel we restrict the discussion to a specific non-linear function $n(x)=x^2$, which reduces the sum over $k$ in Eq.\,(\ref{faa}) to $k_{\text{max}}=2$ as $n^{\prime\prime\prime}=0$.\\
Let us take $m=6$ to set the ideas. The expansion of $u_{f,6}$ involves the Bell polynomials $B_{5,1}$ and $B_{5,2}$ given by
\begin{equation}
\begin{split}
&B_{5,1}(x_1,x_2,x_3,x_4,x_5)=x_5 \\
&B_{5,2}(x_1,x_2,x_3,x_4)=10x_2x_3+5x_1x_4,
\end{split}
\end{equation}
or $B_{m,k}=\sum_{n=1}^{n_{m,k}} b_n$, where $n_{m,k}$ is the number of monomials $b_n$ for the polynomial $B_{m,k}$. In this notation Eq.\,(\ref{faa}) can be written as 
\begin{equation}
\begin{split}
u_{f,6}=&\frac{1}{5!}\int_a^b K(x,t) \sum_{k=1}^{2}\sum_n^{n_{m-1,k}} n^{(k)}(u_0(t)) \times \\ 
 &b_n\left(u_{f,1}(t),\dots,(m-k-1)! u_{f,m-k-1}(t)\right) dt.
 \end{split}
 \label{faarewritten}
\end{equation}
In this form it is evident that all terms in the expansion can be obtained by writing down all possible configurations of $(k,n)$. For example, choosing the configuration $(k=2,n=1)$ yields the term
\begin{equation}
\frac{1}{5!}\int_a^b K(x,t) \, 20 \,  \left( 2! u_{2}(t) \right)  \left( 3! u_{3}(t) \right) dt
\label{preterm}
\end{equation}
by considering the first monomial of $B_{5,2}$ and $n^{\prime\prime}=2$.  We have dropped the subscript $f$ because no ambiguity is possible.
What we achieved so far is just the first step in obtaining a term in the tree expansion since $u_2(t)$ and $u_3(t)$ have to be expanded further. Writing down Eq.\,(\ref{faarewritten}) for $u_2(t)$ and $u_3(t)$ and choosing for each a new configuration $(k,n)$ eliminates $u_2$ and $u_3$ from (\ref{preterm}). To be concrete, choosing $(k=1,n=1)$ for $u_2(t)$ and $(k=2,n=1)$ for $u_3(t)$ yields
\begin{equation}
\begin{split}
&\frac{1}{5!}\int_a^b K(x,t)\, 20 \, \times \\
& 2! \left(  \int_a^b K(t,t^\prime) 2\,u_0(t')\,\, u_1(t^\prime) u_1(t^\prime) dt^\prime \right) \times \\
& 3! \left( \frac{1}{2!} \int_a^b K(t,t^{\prime\prime}) 2 u_1(t^{\prime\prime})\, u_1(t^{\prime\prime}) dt^{\prime\prime}  \right) dt.
\end{split}
\label{exampleterm}
\end{equation}
At this point only $u_1$ remains to be eliminated, for which there is only one possibility,
\begin{equation}
u_1(x)=\int_a^b K(x,t) \, u_0(t) \, u_0(t) dt.
\label{eq:u_1}
\end{equation}
Graphically, this elimination procedure is depicted by rooted trees where the basic elements, see Fig.\,\ref{fig:rootedtreeelements}, are the roots and leafs of the tree which are connected by branches.
One term in the tree expansion corresponds to a fully grown rooted tree. A random term in the tree expansion is picked by growing a random rooted tree in the following way:
\begin{enumerate}
\item Select a random integer $m$ for the root. 
\item Grow a branch from the root according to some random integer $k$, which fixes the Bell polynomial $B_{m-1,k}$. 
\item Grow leafs from this branch according to some random integer $n$, corresponding to the monomials of the Bell polynomials  $B_{m-1,k}$. 
\item Regard every leaf of the branch of $B_{m-1,k}$ as a new root and go back to step 2 if $m > 1$ or finish the recursion by using Eq.~(\ref{eq:u_1}) if $m=1$.
\end{enumerate}
Applied to the example discussed above (see also Fig.\,\ref{fig:construction}), the root is $m=6$ which has two different branch types $k$ where $k=2$ is picked. By selecting $n=2$ two leafs are grown on this branch (cf. Eq.~\ref{exampleterm}), and the decomposition ends by invoking Eq.~(\ref{eq:u_1}).
In order to associate to each fully grown rooted tree a term in the tree expansion each element in the rooted tree must correspond to an element in expression (\ref{exampleterm}) according to the following rules:
\begin{figure}
\centering
	\includegraphics[width=0.45\textwidth]{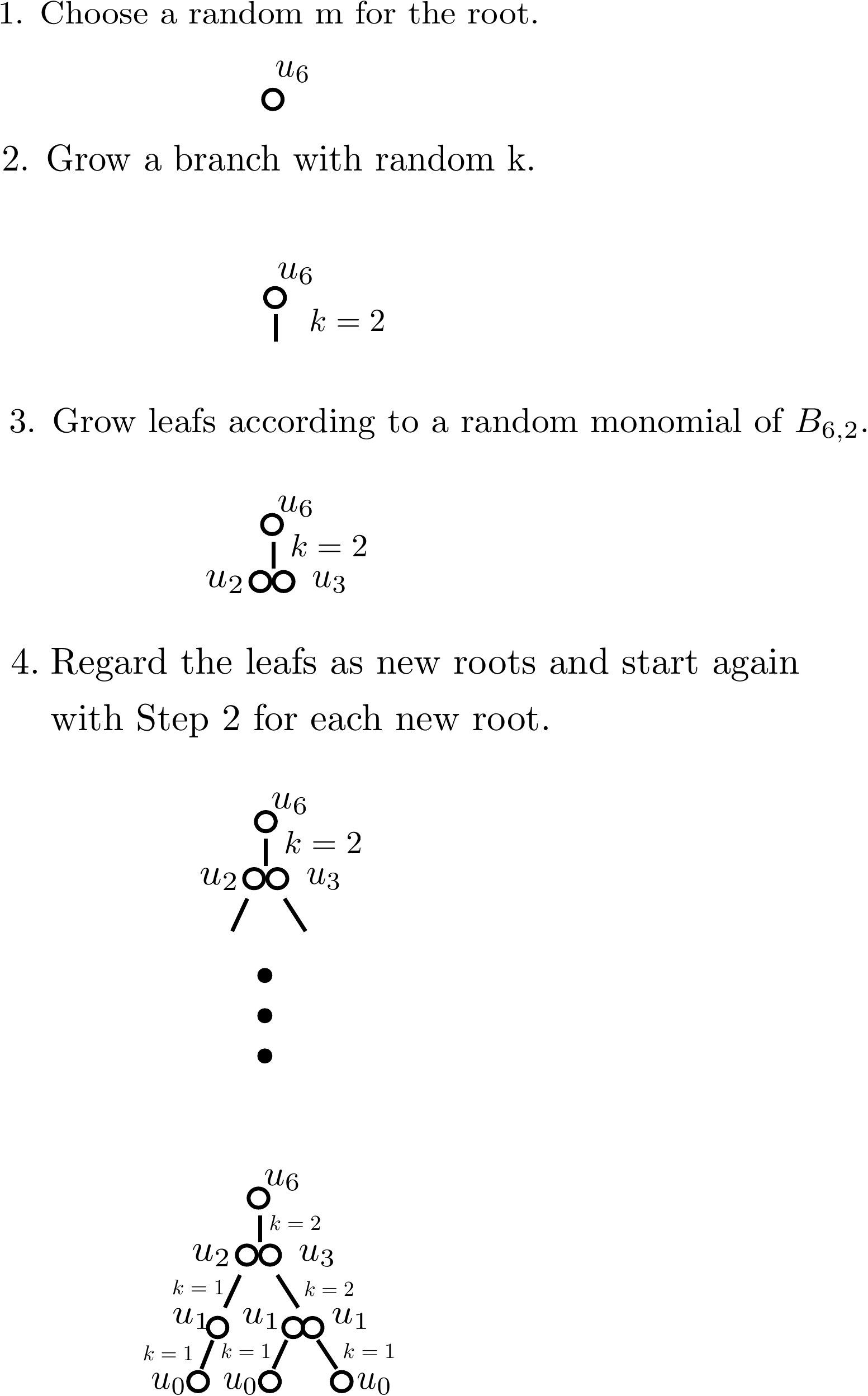}
\caption{The construction of a rooted tree with  height $m=6$. A fully grown rooted tree corresponds to a single term in the tree expansion which is constructed by recursively applying the definition of the root finding, cf. Eq.~(\ref{faa}).}
\label{fig:construction}
\end{figure}
\begin{enumerate}
\item For each branch from a root with given $m$
	\begin{enumerate}
		\item there is a factor $\frac{1}{(m-1)!}$.
		\item there is a factor $2u_0(t)$ if $k=1$ or  $2$ if $k=2$.  
		\item there is the prefactor from the randomly picked monomial of $B_{m-1,k}$.
		\item there is an integration over a new variable $t$ and a factor $K(x,t)$.
	\end{enumerate}
\item For each new leaf with label $m \neq 0$ there is a factor $m!$.
\item For each new leaf with label $m=0$ there is a factor $u_0(t)$.
\end{enumerate}
Fig.\,\ref{fig:fullgrowntree} shows the fully grown, labelled tree, from which the integral can be read off,
\begin{equation}
\begin{split}
\int&\,dt_1 dt_2 dt_3 dt_4 dt_5 dt_6 \, \frac{1}{5!}K(x,t_1) 20 \times \\ 
& \,2! K(t_1,t_3)\,2\,u_0(t_3)\,K(t_3,t_4)\,2\,u_0(t_4)u_0(t_4)\, \times \\ 
& \frac{3!}{2!} K(t_1,t_2)\, 2\, K(t_2,t_5) \, 2\, u_0(t_5) u_0(t_5) \, \times \\
 & K(t_2,t_6) \,2\, u_0(t_6)u_0(t_6).
\end{split}
\label{fullexpression}
\end{equation}
It corresponds to a single term in the tree expansion of the HAM.\\
The approximation of the root to the $M$-th order is given by $\sum_{m=0}^{M}u_{f,m}$ and therefore the tree expansion consists of generating and evaluating all integral expressions corresponding to all possible rooted trees. As the final answer is given by the sum over all deformations $u_{f,m}$ trees with variable height have to be considered.\\
In principle this expansion can be done explicitly by drawing and calculating each diagram. But already for moderate $M$ the exponential growth in the number of diagrams renders this approach impractical. Therefore only a stochastic evaluation of the the sum $\sum_{m=0}^{M}u_{f,m}$ is feasible.
\subsection{Update structure for the Diagrammatic Monte Carlo sampling of rooted trees}
\label{sec:HAM_updates}
\begin{figure}
\centering
\includegraphics[width=0.45\textwidth]{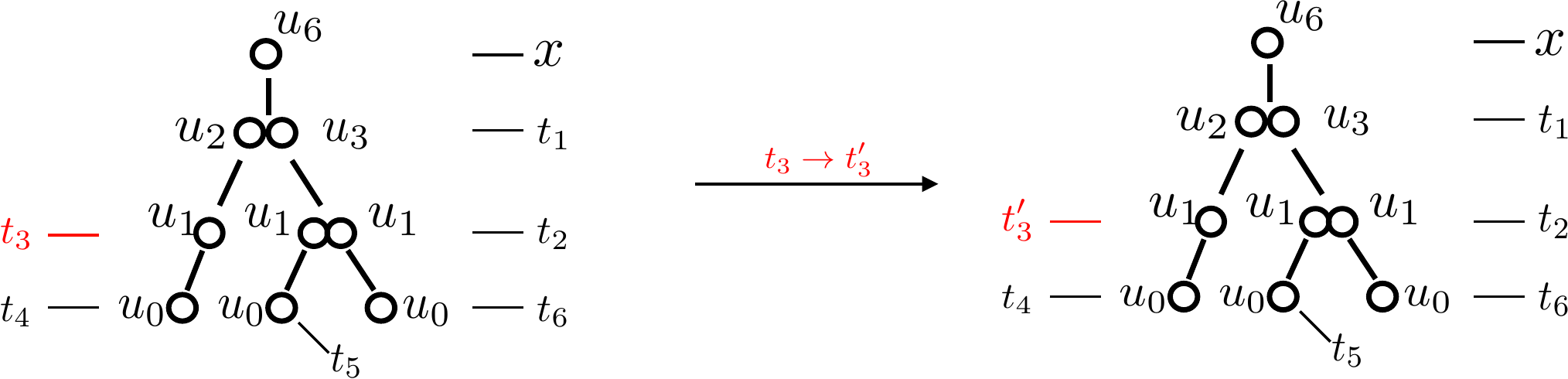}
\caption{The update to change an integration variable. The acceptance ratio is only determined by the different weights of the rooted tree.}
\label{fig:changeintegration}
\end{figure}
\begin{figure}
\centering
\includegraphics[width=0.45\textwidth]{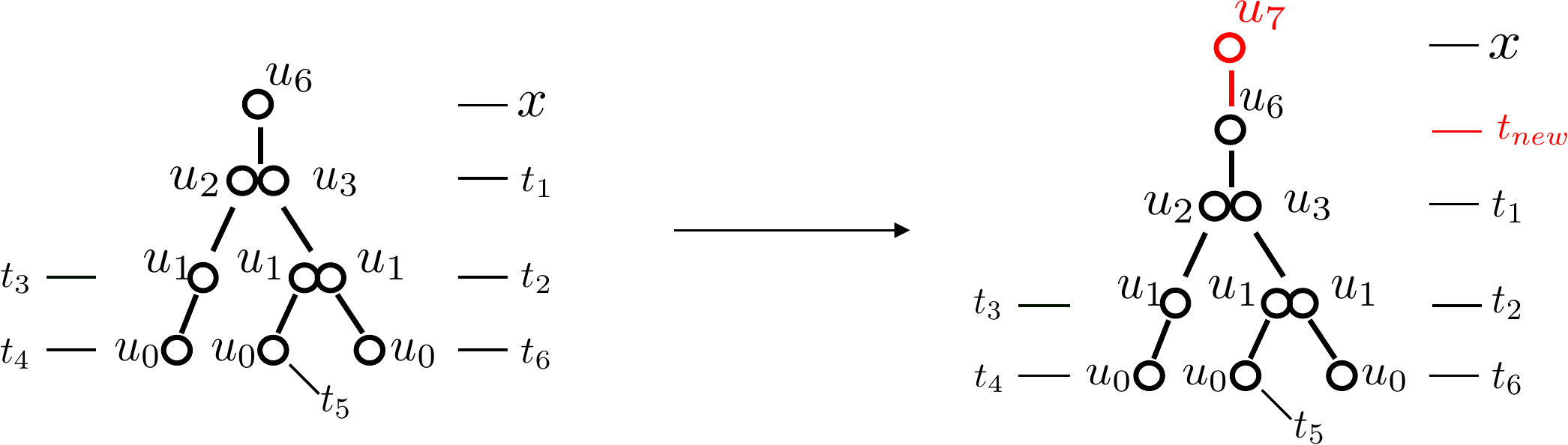}
\caption{The update to change the height of a rooted tree by a branch with $k=1$.}
\label{fig:shrinkexpand}
\end{figure}
\begin{figure}
\centering
\includegraphics[width=0.45\textwidth]{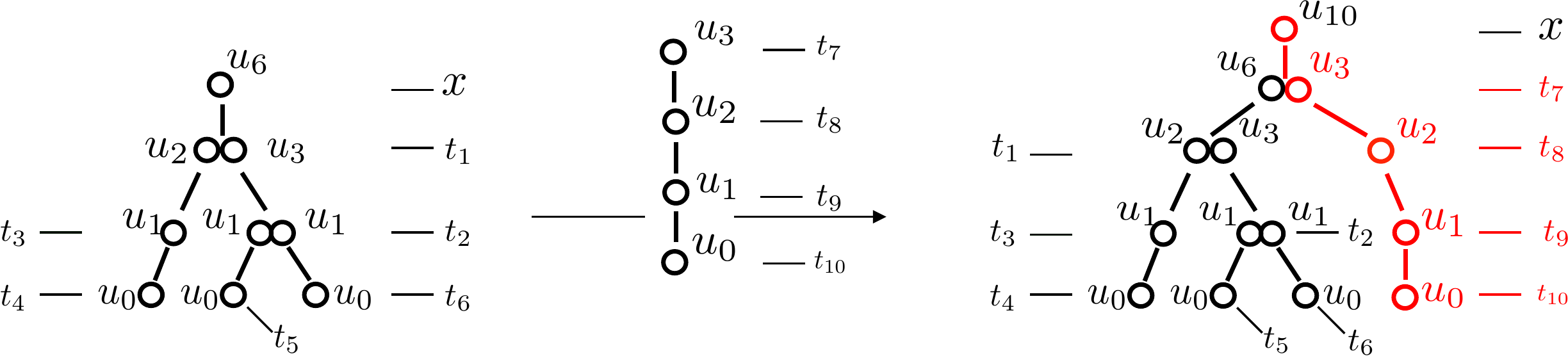}
\caption{The update to change the height of a rooted tree by a branch with $k=2$.}
\label{fig:shrinkexpandcluster}
\end{figure}
\begin{figure}
\centering
\includegraphics[width=0.45\textwidth]{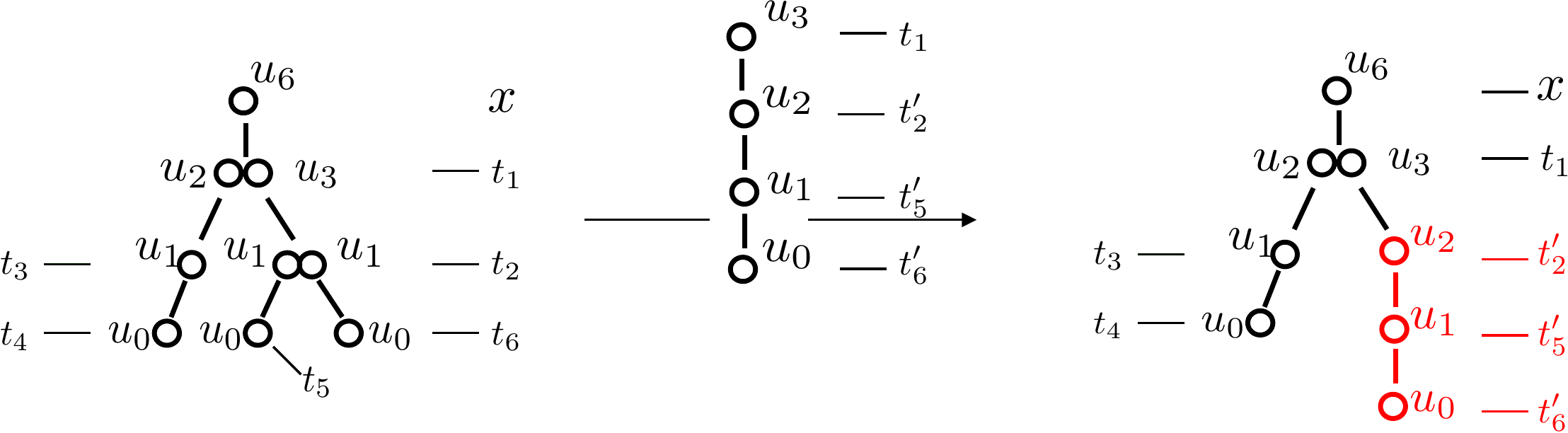}
\caption{The update to change a random subtree of a rooted tree.}
\label{fig:changerandomtree}
\end{figure}
This section introduces the Monte Carlo updates to stochastically sum the tree expansion of the HAM for the above example where $n(x)=x^2$. As in a generic diagMC sampling scheme the core of the algorithm is a Markov Chain which changes the topologies and integration variables of the diagrams according to their respective weights. The weight of a diagram is given by the integration kernel of the integral expression corresponding to the diagram. The rooted tree in Fig.\,\ref{fig:fullgrowntree} corresponds to the integral expression (\ref{fullexpression}) and therefore its weight is
\begin{equation}
\begin{split}
& dt_1 dt_2 dt_3 dt_4 dt_5 dt_6  \, 32\,K(x,t_1)K(t_1,t_3)K(t_3,t_4) \times \\  
& \times  K(t_1,t_2)K(t_2,t_5)K(t_2,t_6)u_0(t_3)u_0(t_4)^2u_0(t_5)^2u_0(t_6)^2. \nonumber
\end{split}
\end{equation}
Furthermore it has been shown that the integral expression can be read off from the rooted tree diagram by using a set of rules which associate to each elementary diagram element, Fig.\,\ref{fig:rootedtreeelements}, an analytic expression. With this set of rules changes in the topology and integration variables of a diagram by introducing or removing elementary diagram elements can be related to changes of the weight of the diagram. In the following a set of updates is introduced which generates a Markov Chain Monte Carlo in the space of all possible rooted trees and therefore stochastically sums the tree expansion of the HAM. The updates to calculate $\sum_{m=0}^{M}u_{f,m}(x)$ for a fixed external variable $x$ are:
\begin{enumerate}
	\item change integration variable: This update performs a shift of the internal integration variables.  It picks one of the internal integration variables and updates the variable according to standard detailed balance rules. As this update is balanced
	with itself and does not change the diagram structure only the weights of the diagrams have to be taken into account. For the above example the update is schematically depicted in Fig.\,\ref{fig:changeintegration}.
	\item shrink-tree/expand-tree: This update is changing the height of the rooted tree.
		\begin{enumerate}
			\item expand-tree: A new root with $m+1$ will be introduced. From this new root a branch with $k=1$ will be grown and the leaf of this branch is the old root. Furthermore a new integration variable $t_{new}$ will be seeded according to some 				probability density function $P(t_{new})$ and assigned to the old root.
			\item shrink-tree: This update can only be performed if the branch growing from the root has $k=1$. The root of the rooted tree is deleted and the leaf of the root is assigned to be the new root. The acceptance ratio is the inverse of the 					expand-tree update.
		\end{enumerate}
	\item shrink-tree-cluster/expand-tree-cluster: This is almost the same update pair as the update shrink-tree/expand-tree. Instead of introducing a new root with a new branch in the $k=1$ configuration a branch in the $k=2$ configuration is grown. As can be
	seen in Fig.\,\ref{fig:shrinkexpandcluster} this can be done by first growing a new random rooted tree with random height and afterwards glue this tree to the root of the current random tree. These two roots can be regarded as the leafs of a $k=2$ branch grown 
	form the new root of the combined rooted tree.
	\item change-subtree: A random leaf of the rooted tree is chosen from which, regarded as a sub-root, a new subtree is grown. When accepted, the old subtree is deleted and replaced by the new subtree.
	This update is shown in Fig.\,\ref{fig:changerandomtree}.
\end{enumerate}
\subsection{Illustration}
\label{sec:HAM_example}
\begin{figure}[t]
\centering
\includegraphics[width=0.45\textwidth]{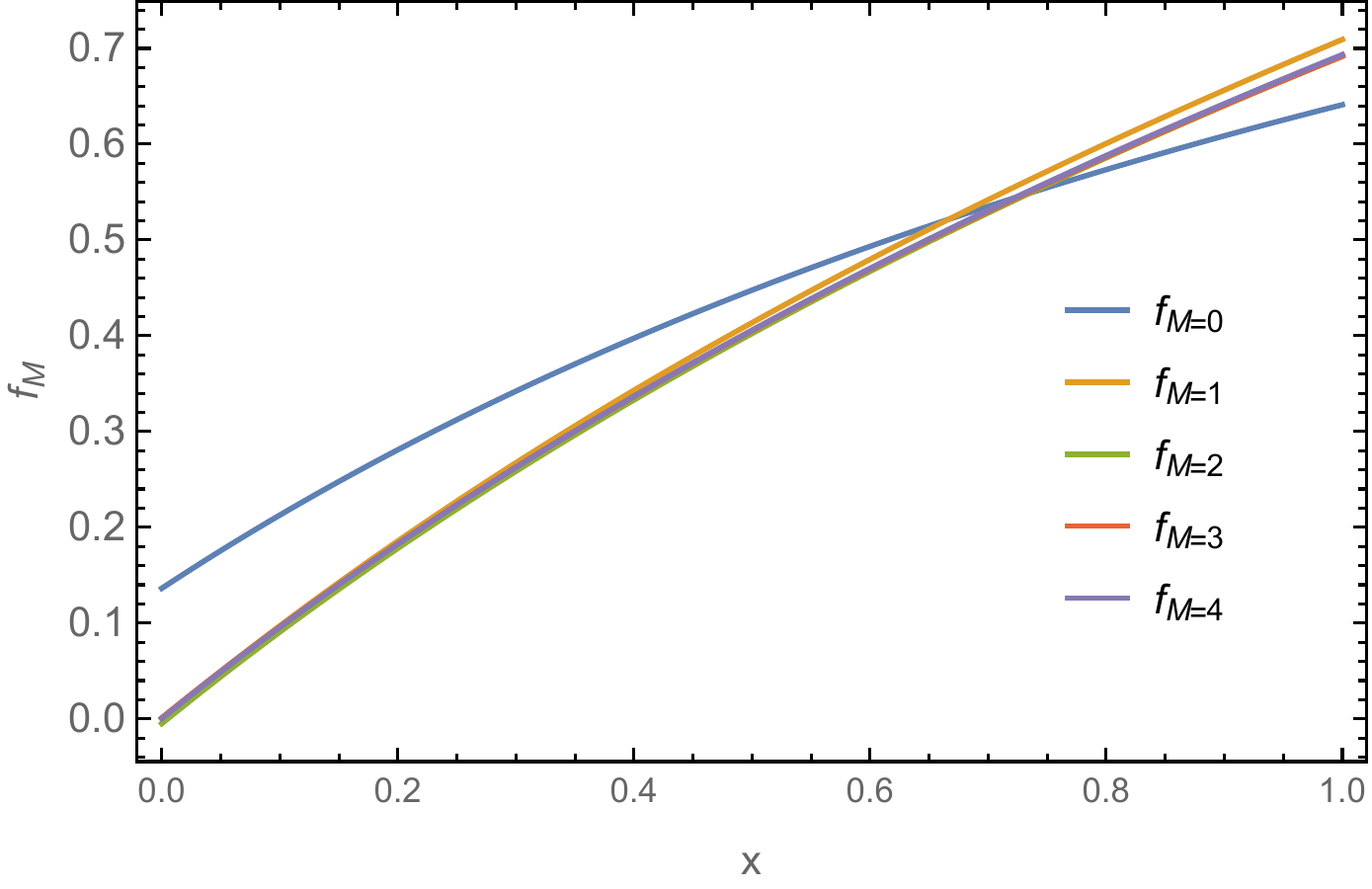}
\caption{The successive approximations $f_M(x) = \sum_{m=0}^{M} u_m(x)$ to the solution of the example integral equation Eq.~(\ref{exampleintegral}). The initial guess $u_0=f_0$ for the solution of the integral equation converges to the correct solution already for low M $\approx$ 4 iteration steps in the HAM.}
\label{HAMresults}
\end{figure}
\begin{table}[b]
\centering
\resizebox{0.45 \textwidth}{!} 
{
\begin{tabular}{|l|l|l|l|}
\hline
\multicolumn{1}{|c|}{m} & $u_m(x=0.5)$ & $u_m(x=0.5)$ (diagMC)  & p   \\ \hline
                    1   & -0.0339665 &  &    \\ \hline
                    2  & -0.0112517 &  -0.01127  $\pm$ 0.00004  &  0.587  \\ \hline
                    3   & 0.00292815  & 0.00293 $\pm$ 0.00001 &  0.247  \\ \hline
                    4   & 0.000486724 & 0.000488 $\pm$ 5e-06 & 0.096    \\ \hline
                    5   & -0.000359147 & -0.000359 $\pm$ 5e-06 & 0.041   \\ \hline
                    6   & -2.77262e-06 & -2.6 e-06 $\pm$ 3.6e-06 &  0.019   \\ \hline
                    7   &  4.24149e-05 & 4.2e-05 $\pm$ 3e-06 &  0.009  \\ \hline
\end{tabular}
}
\caption{Comparing the tree expansion sampled with diagMC to the exact answer where every $u_{m}(x)$ is calculated and stored on an external $x \in [0,1]$ grid. The last column indicates the probability $p$ of reaching the expansion order $m$.}
\label{table:diagMCresults}
\end{table}
We illustrate the above algorithm for a one-dimensional integral equation,
\begin{equation}
f(x)=c(x)+\int^{1}_{0}K(x,t)n(f(t))dt.
\label{exampleintegral}
\end{equation}
The kernel of the integration is $K(x,t)=(x-t)$ and $n(x)=x^2$. The function $c(x)$ is picked in such a way that the solution of Eq.~(\ref{exampleintegral}) is given by $f(x)=\log(x+1)$. It can be checked that $c(x)=\log(x+1)+2 \log(2)-2x(\log(2)-1)(\log(2)-1)-\frac{5}{4}$ satisfies this condition. The $m$-th order deformation equation is given by
\begin{equation}
u_m(x)=\int_0^1 dt \, K(x,t)\left(\sum_{k=0}^{m-1}u_k(t)u_{m-k-1}(t)\right).
\end{equation}
The initial approximation of the root is $u_0(x)=c(x)$ and $h=H(x)=1$. As is shown in Fig.\,\ref{HAMresults} the result $f(x) \approx f_M(x) = \sum_{m=0}^{M} u_m(x)$ quickly converges to the correct answer as a function of maximum deformation order $M$. 
The results of the diagMC root finding calculations for $x=0.5$ are given in Table~\ref{table:diagMCresults} and compared with the exact answer. As the expansion of $u_1$ into $u_0$ is immediate it has not been included into the diagMC procedure. \\
The 4-th column in Table~\ref{table:diagMCresults} shows the probability $p$ to reach deformation order $m$, indicating the convergence of the diagMC root finding algorithm as $p \to 0$ for larger $m$.\\
In the next section we extend the proof of principle for the HAM method to the more challenging case of the DSEs for $\phi^4$ field theory.
%
%
%
%
\begin{figure}[t]
\centering
\includegraphics[width=0.45\textwidth]{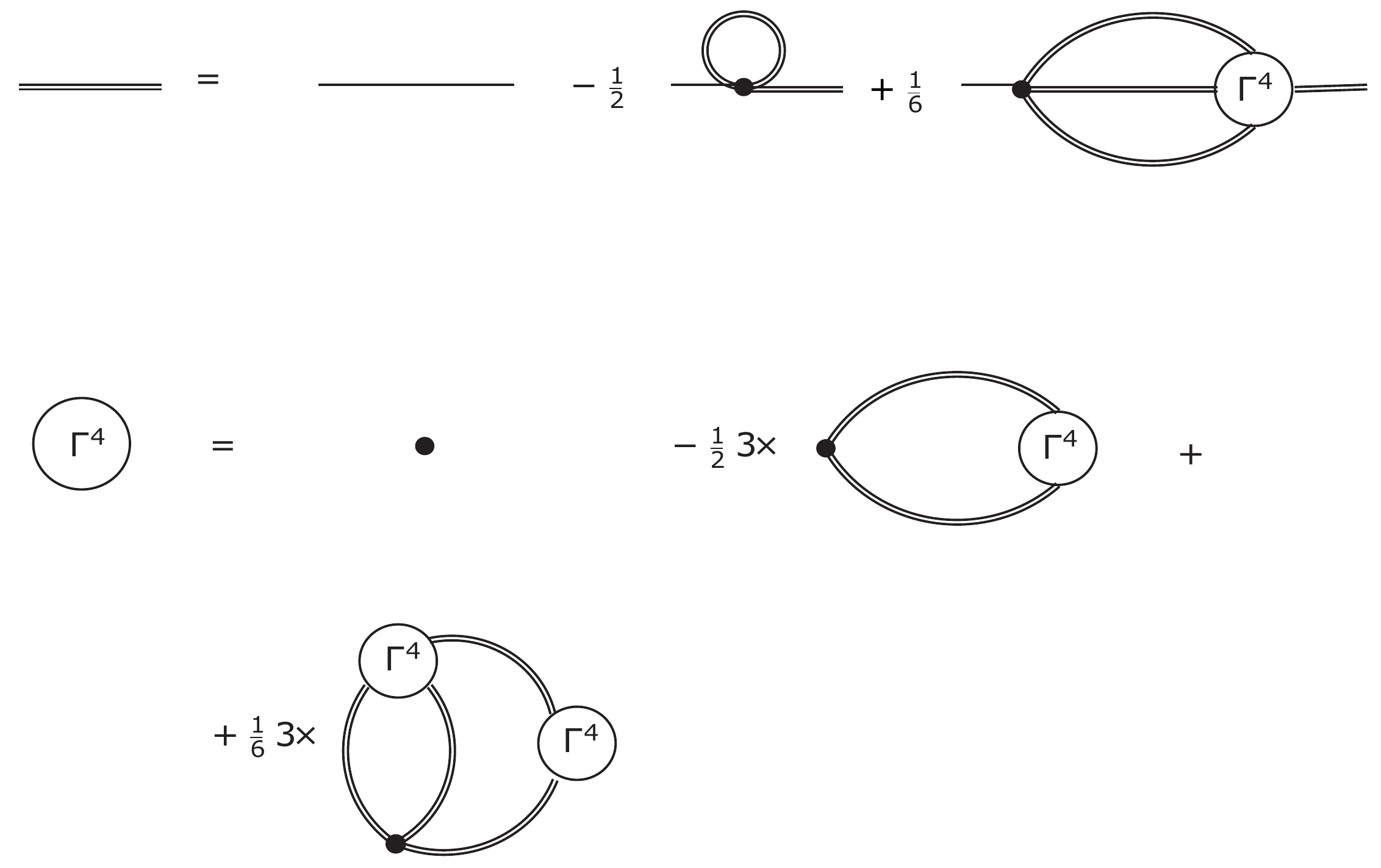}
\caption{The diagrammatic representation of the truncated tower of DSE for the $\phi^4$ theory. The permutation of external legs is assumed implicitly and indicated by $3 \times$. }
\label{fig:DSEphi4}
\end{figure}
\section{Solving Dyson-Schwinger Equations for the One-Dimensional $\phi^4$ Theory}
\label{sec:four}
Since it is not a priori clear if (and how) the HAM equations and the stochastic sampling of rooted trees work in practice for the coupled set of DSEs, we analyze the HAM method for the DSEs of the 1D $\phi^4$ theory in two steps: First, in Sec.~\ref{sec:phi4HAM}, we use the HAM for ``Solver I" and ``Solver II" separately, cf. Fig.\;\ref{fig:selfconsistencyloop}. This is possible because the storage of the vertex function poses no problem in 1D. We demonstrate superior convergence properties compared to fixed point iterations.  Second, in Sec.~\ref{sec:phi4_Gamma}, we check the stochastic evaluation of the tree expansion up to 8th order (given the exact 2-point correlation function) against the result obtained in the first step (Sec.~\ref{sec:phi4HAM}). These are preliminary steps for the stochastic solution of the coupled set of DSEs in 2D including both 2-point and 4-point functions and which requires the combination of both steps, and which will be presented in Sec.~\ref{sec:five}. \\
\\
\subsection{Model and notation}
The truncated set of DSEs for the $\phi^4$ theory \cite{PelsterGlaum} is in the thermodynamic limit given by
\begin{widetext}
\begin{equation}
\begin{split}
&G_{1,2}=G_{0;1,2}+\int_{3,4}G_{0;1,3}\Sigma_{3,4}G_{4,2} \\
&\text{where} \hspace{5mm} \Sigma_{1,2} = - \frac{\lambda}{2} \delta(1-2)G_{1,1}+\frac{\lambda}{6}\int_{3,4,5}G_{1,3}G_{1,4}G_{1,5}\Gamma_{3,4,5,2} \\
&\Gamma_{1,2,3,4} = \lambda \delta(1-2)\delta(1-3)\delta(1-4) -\frac{\lambda}{2}\int_{5,6}G_{1,5}G_{1,6}\left[\Gamma_{5,6,3,4}+\Gamma_{5,6,2,4}+\Gamma_{5,6,3,2}\right]+ \\
& \hspace{15mm} \frac{\lambda}{6}\int_{5,6,7,8,9}G_{1,5}G_{1,6}G_{7,8}G_{1,9} \left[\Gamma_{5,2,7,6}\Gamma_{9,8,3,4}+\Gamma_{5,3,7,6}\Gamma_{9,8,2,4}+\Gamma_{5,4,7,6}\Gamma_{9,8,3,2}\right]\, .
\end{split}
\label{truncatedPhi4DSE}
\end{equation}
\end{widetext}
For later purposes (cf. Sec.~\ref{sec:five}) we leave the dimensionality $D$ of the lattice, the bare mass $m^2$, and the bare coupling $\lambda$ as free parameters. The lattice constant is fixed, $a=1$ and the system size infinite unless otherwise specified. In the above equations the sum runs over all possible lattice points $\mathbf{r_i}$ and is denoted by $\int_{i}$. $G_{1,2}$ denotes the 2-point correlation function $G(\mathbf{r_1},\mathbf{r_2})$ and $\Gamma_{1,2,3,4}=\Gamma(\mathbf{r_1},\mathbf{r_2},\mathbf{r_3},\mathbf{r_4})$ denotes the 4-point vertex function. Terms of order $\mathcal{O}(\Gamma^6)$ are neglected.
$G_{0;1,2}$ is the 2-point correlation function of the Gaussian model ($\lambda$=0) given by
\begin{equation}
G_0(\mathbf{r})=\int \frac{d^D\mathbf{p}}{(2\pi)^D} e^{-i \mathbf{p} \mathbf{r}} \left(4\sum_{i=1}^{D}\text{sin}^2(\frac{p_i}{2})+m^2\right)^{-1} \, . \nonumber
\end{equation}
The truncated DSEs, diagrammatically shown in Fig.\,\ref{fig:DSEphi4}, are regarded as root finding problems and solved by the HAM.
The $m$-th order deformations of $G$ and $\Gamma$ are denoted by $u_{G,m}$ and $u_{\Gamma,m}$, respectively, and are given by
\begin{widetext}
\begin{eqnarray}
u_{\Gamma,m}(\mathbf{x}) & = & \chi_mu_{\Gamma,m-1}(\mathbf{x}) - h \left[   u_{\Gamma,m-1}(\mathbf{x}) - \lambda\delta(\mathbf{x}) \tilde{\chi}_m - \frac{\lambda}{2}\sum_c\int_{5,6} K_c(\mathbf{x},5,6)u_{\Gamma,m-1}(f_c(\mathbf{x},5,6)) +  \right. \nonumber \\ 
{} & {} &\left. \frac{\lambda}{6}\sum_c\int_{5,6,7,8,9}K_c(\mathbf{x},5,6,7,8,9) \sum_{k=0}^{m-1} u_{\Gamma,k}(f_c(\mathbf{x},8,9))u_{\Gamma,m-1-k}(f_c(\mathbf{x},5,6,7))    \right].  \label{deformingGamma} \\ 
u_{G,m}(1-2) & = & \chi_m u_{G,m-1}(1-2) - h \left[ u_{G,m-1}(1-2)+\tilde{\chi}G_0(1-2)  \right. \nonumber \\ 
{} & {} & - \frac{\lambda}{2}\sum_{k=0}^{m-1}\int_3 G_0(1-3)u_{G,m}(0)u_{G,m}(3-2) \nonumber \\ 
{} & {} & + \frac{\lambda}{6}\sum_{k=0}^{m-1}\sum_{i=0}^{k}\sum_{j=0}^{m-1-k}\int_{3-7}u_{G,i}(3-5)u_{G,k-i}(3-6) \times \nonumber \\
{} & {} & \left. u_{G,j}(3-7)u_{G,m-1-k-j}(4-2) G_0(1-3)\Gamma(7-6,7-5,7-4) \right].   \label{deformingG}
\end{eqnarray}
\end{widetext}
The linear operator $\mathcal{L}$ (cf. Eq.~(\ref{eq:HAM_sys})) is chosen to be the identity operator. \\
\begin{figure}[t]
\includegraphics[width=0.45\textwidth]{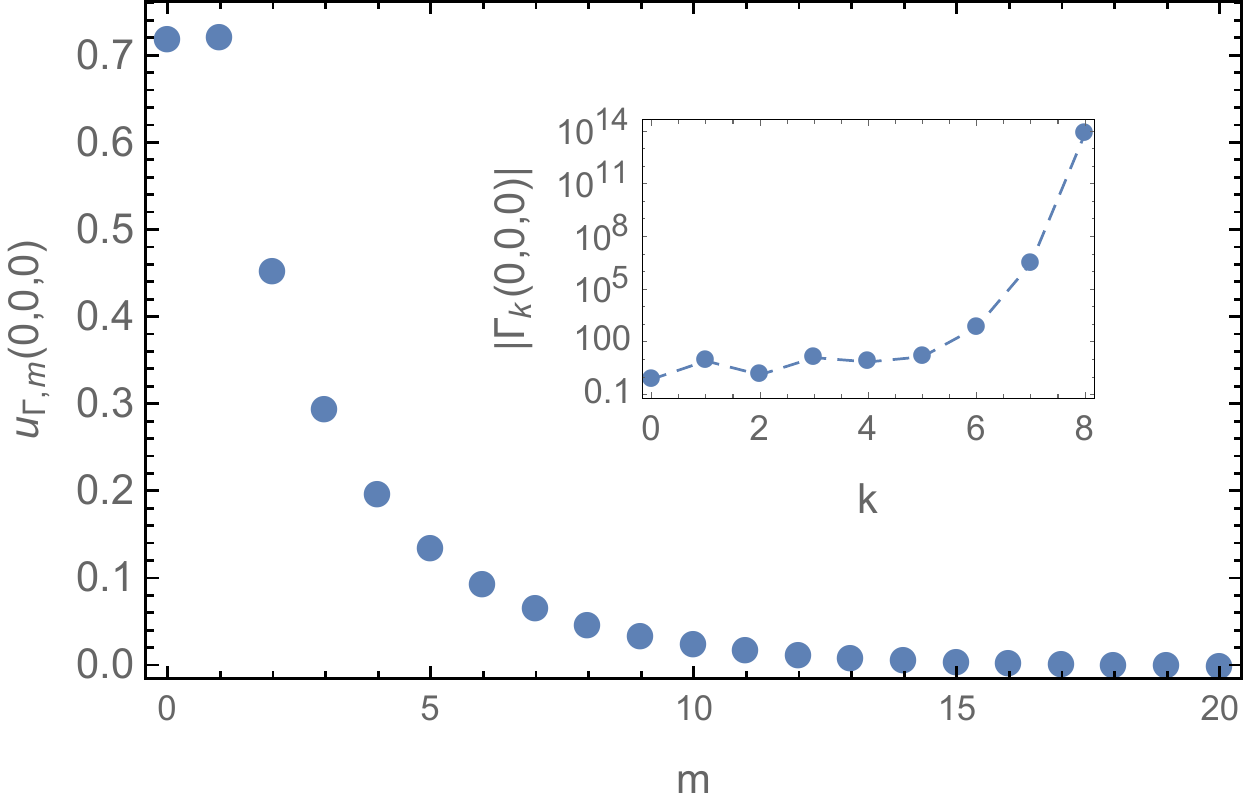}
\caption{The convergence of the deformations $u_{\Gamma,m}$ at $(s,t,u)=(0,0,0)$. The inset shows the divergence of a fixed point iteration for the solution of the DSE of the vertex function. Here $k$ is the number of fixed point iterations.}
\label{fig:convergenceGamma}
\end{figure}
The notation is as follows:
\begin{enumerate}
\item The external coordinates are denoted as $\mathbf{x}=(\mathbf{s},\mathbf{t},\mathbf{u})$, where due to translational invariance $(\mathbf{s},\mathbf{t},\mathbf{u})=(\mathbf{r}_1-\mathbf{r}_2,\mathbf{r}_1-\mathbf{r}_3,\mathbf{r}_1-\mathbf{r}_4)$. They are $D$-dimensional vectors.
\item The sum $\sum_c$ runs over the three possible symmetry channels $c = (\mathbf{s}, \mathbf{t}, \mathbf{u})$ which corresponds to the possible permutations of the external legs, cf. Eq.\,(\ref{truncatedPhi4DSE}).
\item The kernel functions $K_c$ denote the contribution of the 2-point correlation functions to the vertex function. For example, $K_{c=s}(\mathbf{x},5,6)=G(\mathbf{r}_1-\mathbf{r}_5)G(\mathbf{r}_1-\mathbf{r}_6)$ and $K_{c=t}(\mathbf{x},5,6,7,8,9)=G(\mathbf{r}_1-\mathbf{r}_5)G(\mathbf{r}_1-\mathbf{r}_6)G(\mathbf{r}_7-\mathbf{r}_8)G(\mathbf{r}_1-\mathbf{r}_9)$.
\item The functions $f_c$ denote the linear combinations of positions at which the vertex is evaluated, e.g. $f_{c=s}(\mathbf{x},5,6)=(\mathbf{r}_5-\mathbf{r}_6,\mathbf{r}_5-\mathbf{r}_3,\mathbf{r}_5-\mathbf{r}_4)$  and $f_{c=t}(\mathbf{x},5,6,7)=(\mathbf{r}_5-\mathbf{r}_3,\mathbf{r}_5-\mathbf{r}_7,\mathbf{r}_5-\mathbf{r}_6)$.
\item $\chi_m=1$ for $m>1$ and 0 for $m=1$ and $\tilde{\chi}_m=1-\chi_m$.
\item The parameter $h$ is the convergence control parameter of the HAM (cf. Eq.~(\ref{eq:HAM_sys})).
\end{enumerate}
With these two equations it is possible to apply the self-consistency loop of Fig.\,\ref{fig:selfconsistencyloop}. In the case of the 1D $\phi^4$ model the vertex depends only on three variables and therefore the vertex can still be stored  and the HAM root finding of ``Solver I" and ``Solver II" in Fig.\,\ref{fig:selfconsistencyloop} can be implemented separately. 
\begin{figure}[b]
\includegraphics[width=0.45\textwidth]{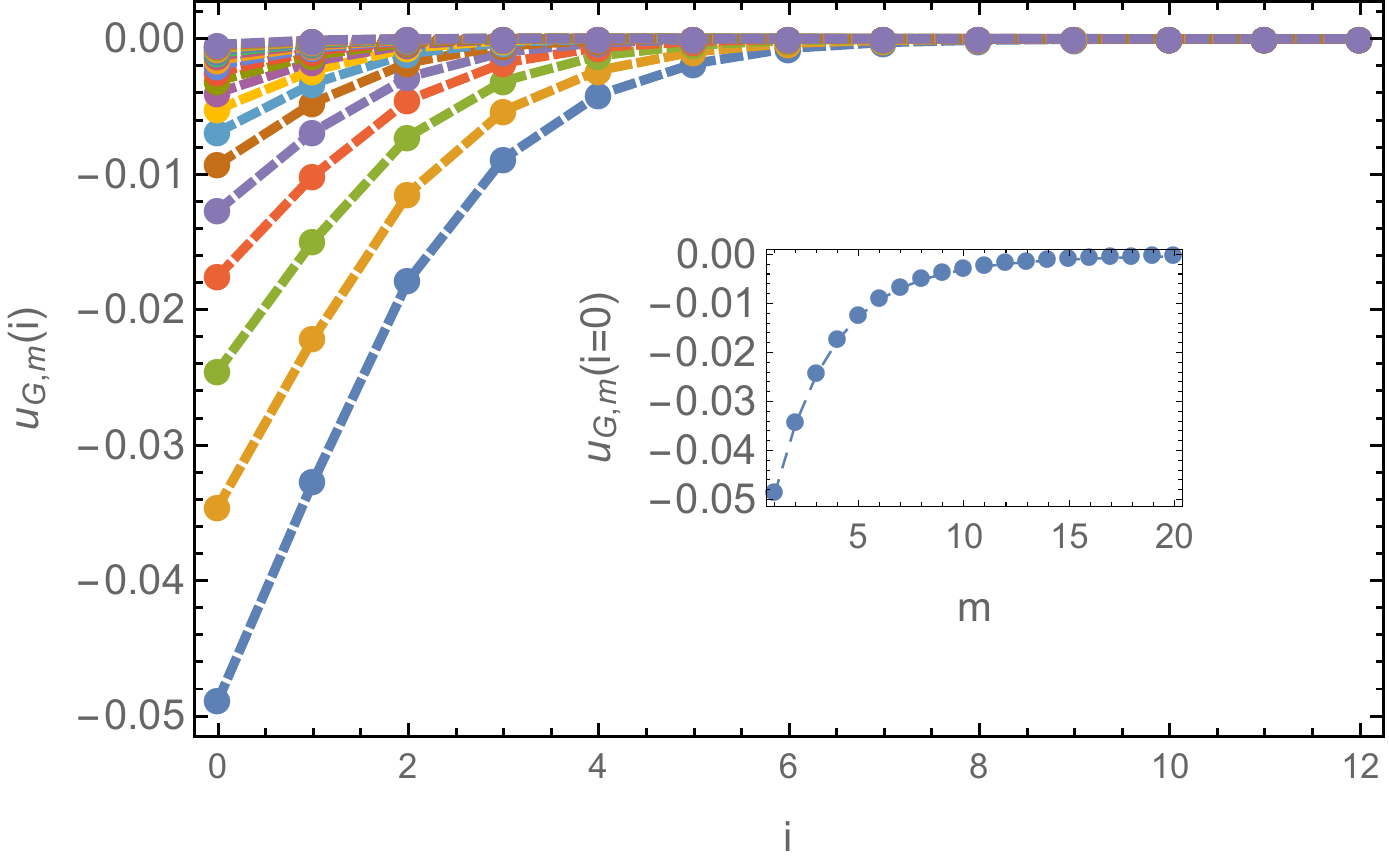}
\caption{Contributions of the $m$-th order deformation $u_{G,m}(i=|r_1-r_2|)$ for increasing order $m$, bottom (lowest value of $m$) to top (highest value of $m$). The convergence at the origin $i=0$ is shown in the inset.}
\label{fig:convergenceG}
\end{figure}
\subsection{Root finding without tree expansion}
\label{sec:phi4HAM}
\begin{figure}[bt]
\includegraphics[width=0.45\textwidth]{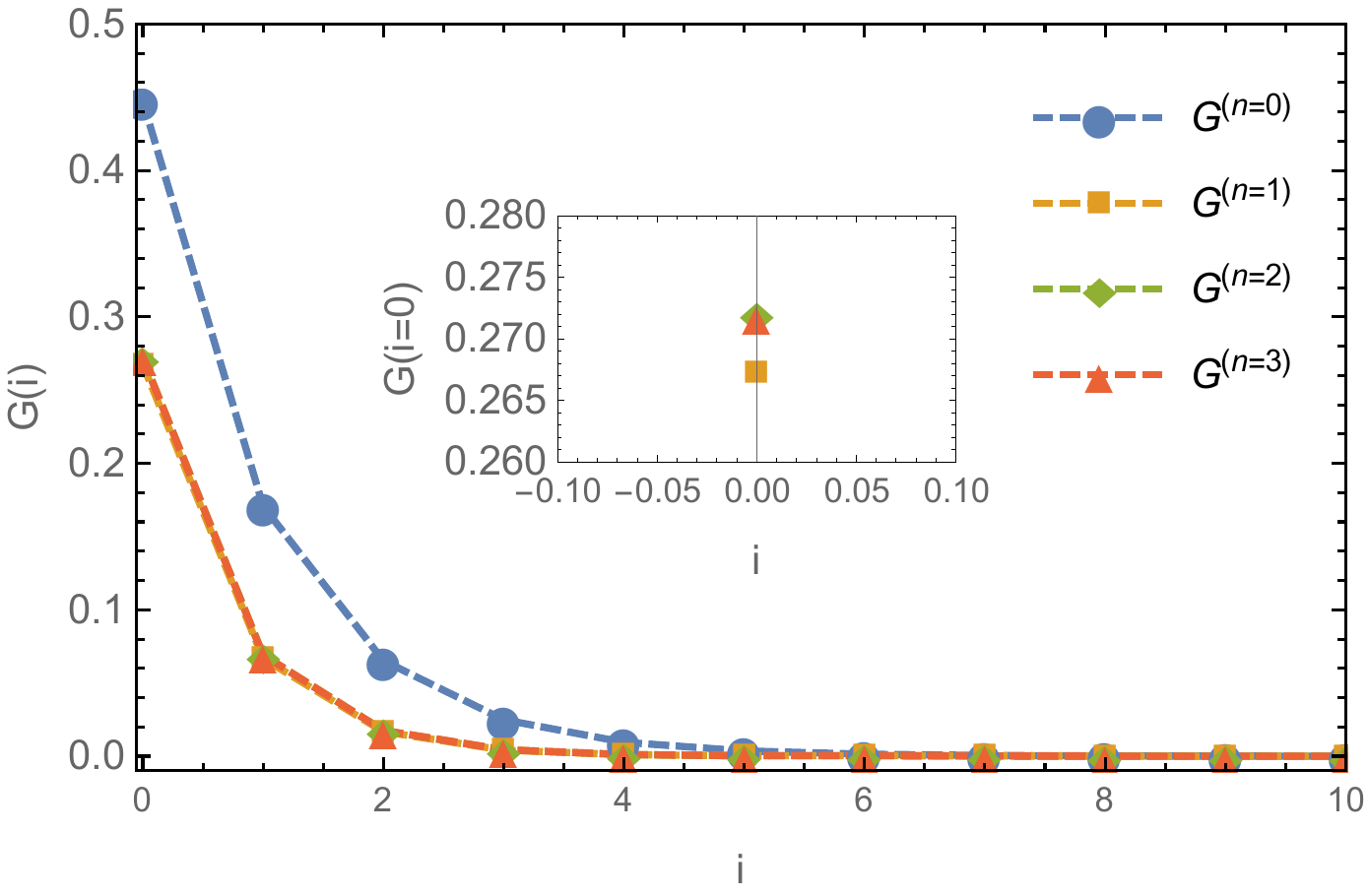}
\caption{The approximations for the full 2-point correlation function $G(i=|r_1-r_2|)$ after $n=0,\dots,3$ self-consistency steps. The inset shows in detail the fast convergence of the self-consistency loop at $i=|r_1-r_2|$.}
\label{fig:convergenceGuessing_2}
\end{figure}

All results presented in this section are for the $D=1$ case where the bare mass and bare coupling are fixed to $m^2=1$ and  $\lambda=10$, respectively. The HAM control parameter is taken as $h=\frac{1}{\lambda}$.
The self-consistency loop is applied in the following way (cf. Fig.\,\ref{fig:selfconsistencyloop}):
\begin{enumerate}
\item In step $n=0$ we start from the guess $G^{(n=0)}=G_0$. 
\item With $G^{(n-1/2)}$ ``Solver I" is searching for the root of the DSE for the vertex by using Eq.~(\ref{deformingGamma}) with starting value the second order perturbative result for the vertex. Let the result be $\Gamma^{(n)}$. 
\item With $\Gamma^{(n)}$ ``Solver II" calculates the solution of the first DSE by using Eq.~(\ref{deformingG}) with starting value $u_{G,0}=G_0$. Let the result be $G^{(n+1/2)}$.
\end{enumerate}
Let us now discuss the results. Fig.\;\ref{fig:convergenceGamma} and Fig.\;\ref{fig:convergenceG} show the convergence of the HAM root finding of ``Solver I" and ``Solver II" for the first loop of the self-consistency, $n=1$. 
The convergence of ``Solver I" is demonstrated at a single point $\Gamma^{(n=1)}(0,0,0)$ but holds for any point $(s,t,u)$. The inset shows the divergence of a fixed point iteration which tries to find the root by brute force iteration.
Fig.\;\ref{fig:convergenceGuessing_2} shows the convergence of the self-consistency loop in $G$ as a function of the loop index $n$. As can be seen in more detail in the inset, the self-consistency already converges at $n=3$. Convergence of the 2-point correlation function $G$ in the self-consistency loop implies convergence of $\Gamma$.
\begin{figure}[t]
\includegraphics[width=0.45\textwidth]{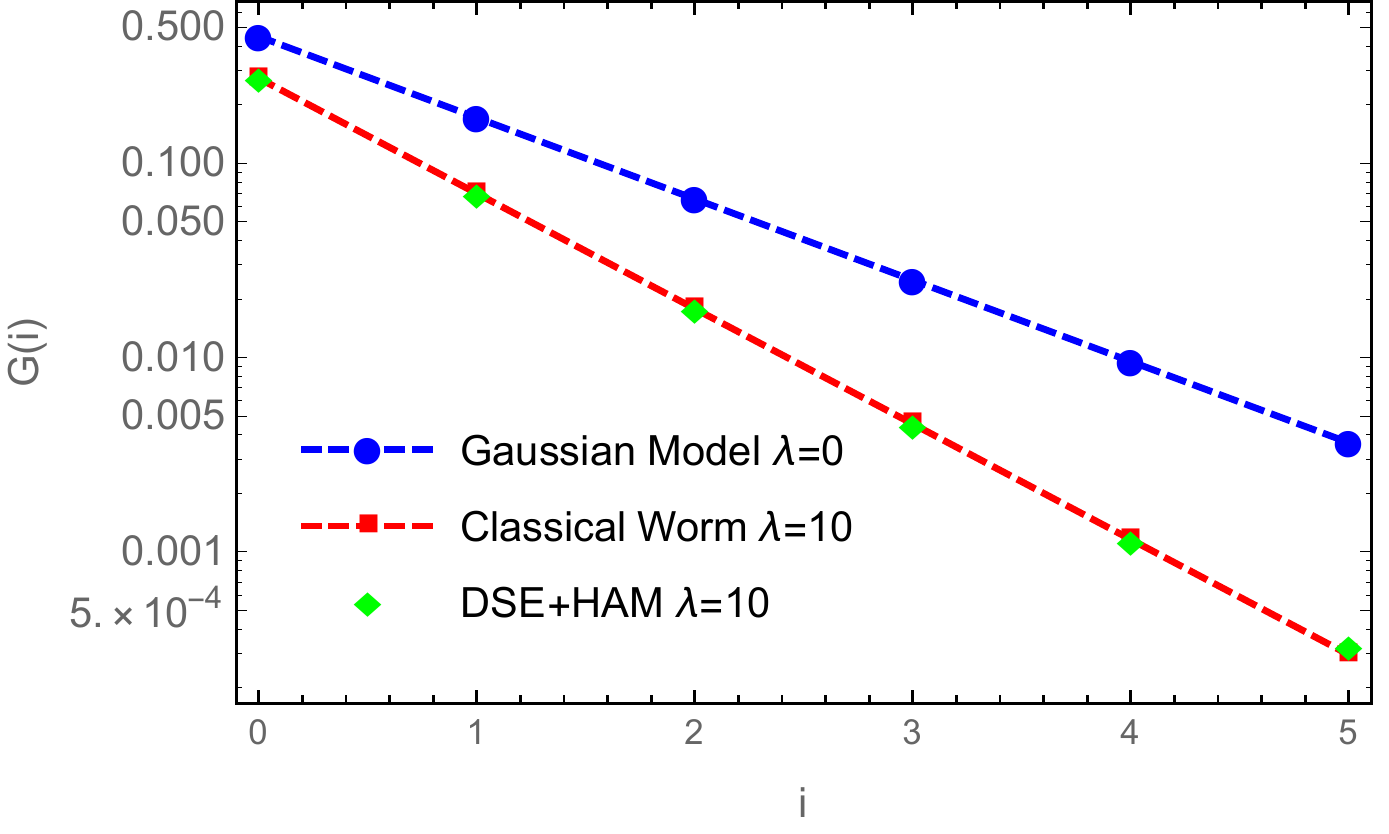}
\caption{The result for the 2-point correlation function $G(i=|r_1-r_2|)$ from the self-consistency loop (HAM+DSE) and from the classical worm algorithm \cite{ClassicalWorm}.}
\label{fig:compare}
\end{figure}
Finally, Fig.\;\ref{fig:compare} shows the comparison of the 2-point correlation function with the one obtained by the classical worm algorithm~\cite{ClassicalWorm}, proving that the self-consistency loop converges to the correct answer.\\
\subsection{Root finding with tree expansion}
\label{sec:phi4_Gamma}
\begin{figure}
\centering
\includegraphics[width=0.45\textwidth]{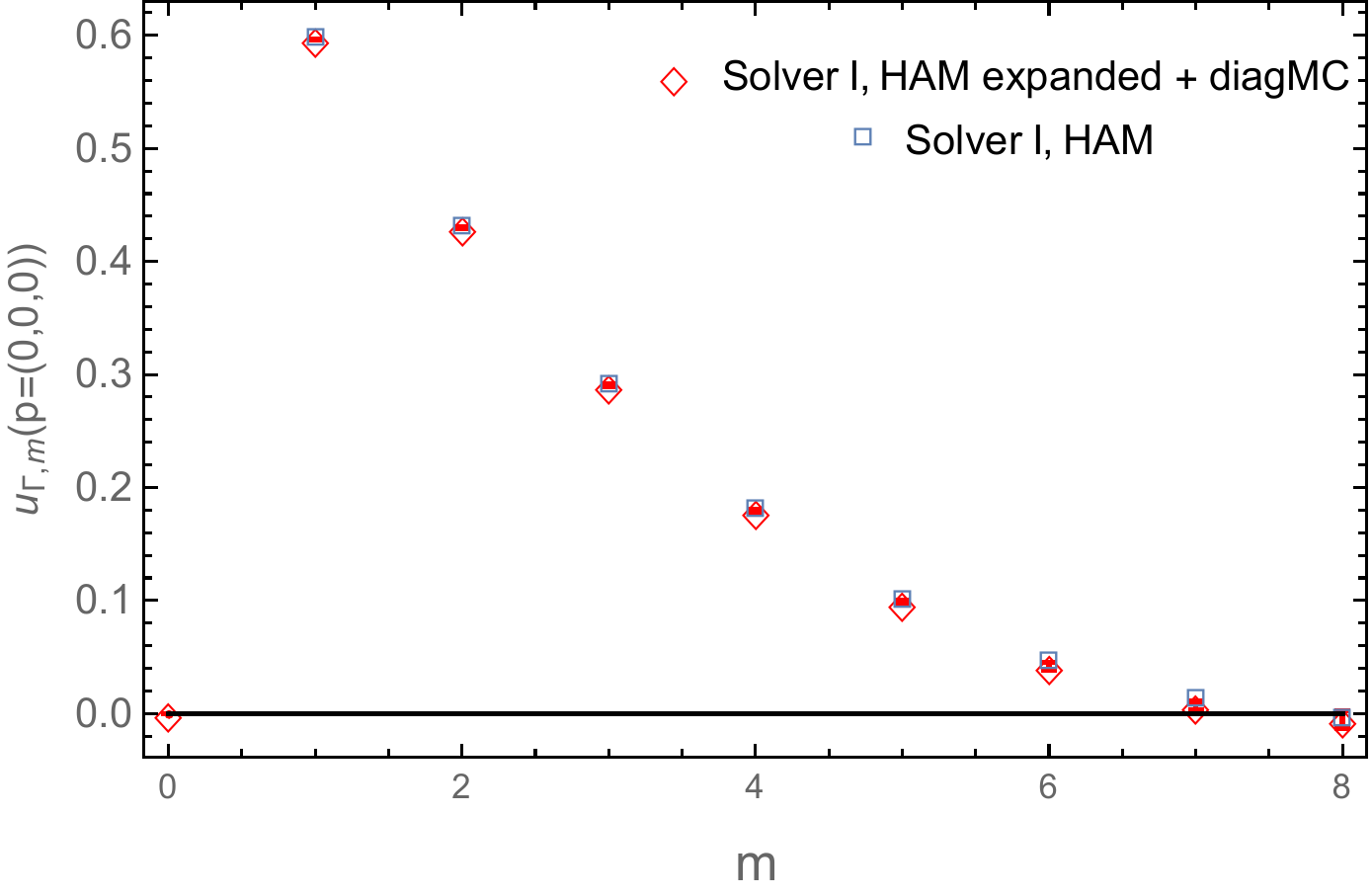}
\caption{Comparing the diagMC sampling of the tree expansion with the results of Sec.\,\ref{sec:phi4HAM} where all $u_{\Gamma,m}$ were calculated and stored in real space. The open squares are the Fourier transforms of the functions $u_{\Gamma,m}$ at zero external momentum.}
\label{fig:convergenceGamma_expanded}
\end{figure}

The goal of this section is to show that the vertex function $\Gamma$ can be obtained by the stochastic evaluation of the tree expansion, i.e. only the DSE for the vertex function is solved with the 2-point correlation function fixed to be the result of the classical worm algorithm. We leave the discussion of the solution for the coupled set to the next section where the DSEs are considered for the 2D setup.\\
In the following the DSE for the vertex function is considered in momentum space where the $m$-th order deformation equation for $\Gamma(\mathbf{p}_1, \mathbf{p}_2, \mathbf{p}_3)$ is given by the Fourier transform of (\ref{deformingGamma}). This is again the starting point for the tree expansion which eliminates all references to deformations $u_{\Gamma,i}$, $i<m$ from the $m$-th order deformation equation.
Statistics for the tree expansion  $\Gamma(0, 0, 0) = \sum_m u_{\Gamma,m}(0, 0, 0)$ is obtained from the estimator
\begin{equation}
\langle u_{\Gamma,m}(0, 0, 0) \rangle_{\text{MC}} = \frac{ \langle \delta_{m,m( \nu_{\text{tree}} )} \rangle_{\text{MC} } }{ \langle \delta_{ \nu_{\text{tree}}, \nu_{N_{\text{tree}}} } \rangle_{ \text{MC} } } N_{\text{tree}},
\end{equation}
where $m( \nu_{\text{tree}} )$ is the current height of the rooted tree diagram.
The configuration space $\{ \nu_{\text{tree}} \}$ of the Monte Carlo (MC) sampling includes rooted tree diagrams with a fixed external momentum configuration, $\mathbf{p}_1 = \mathbf{p}_2 = \mathbf{p}_3 = 0$. $\nu_{N_{\text{tree}}}$ is a normalization diagram whose numerical value $N_{\text{tree}}$ is calculated with a deterministic integration algorithm (in practice we take one of the leading terms in the tree expansion).
\\
The tree expansion of (\ref{deformingGamma}) suffers from an alternating sign originating from the different signs of the linear and quadratic term with respect to $\Gamma$ in the DSE for the vertex function, cf. Eq~(\ref{truncatedPhi4DSE}). For this reason it is impossible to sample the tree expansion to arbitrary large orders. We will discuss the numerical sign problem for our algorithm in more detail in the next section.\\
The results from the sampling of the tree expansion are compared with the straightforward implementation of the HAM algorithm in Sec.\,\ref{sec:phi4HAM} where $\Gamma = \sum_m u_{\Gamma,m}$ was calculated in real space by storing an manipulating the high-dimensional objects $u_{\Gamma,m}$. Fig.\,\ref{fig:convergenceGamma_expanded} compares the results from the tree expansion at $(\mathbf{p}_1, \mathbf{p}_2, \mathbf{p}_3)=(0,0,0)$ with the Fourier transforms of the explicitly stored $u_{\Gamma,m}$. It clearly shows that it is possible to calculate the tree expansion for the case of the 1D $\phi^4$ model up to 8 iteration steps with high accuracy.\\
The tree expansion is a convergent expansion as long as the root finding algorithm is powerful enough to find the solution of the DSE. For the problem under consideration the skeleton series expansion of the Luttinger-Ward functional is asymptotic and therefore breaks down as soon as $\lambda \sim \mathcal{O}(1)$. Fig.\,\ref{fig:convergenceGamma_expanded} shows that the tree expansion considerably increases this parameter regime and seems to be only limited by the sign problem and/or a phase transition (cf. the next section).
\section{Solving Dyson-Schwinger Equations for the Two-Dimensional $\phi^4$ Theory}
\label{sec:five}
\subsection{Problem Setup}
In this section we will move on to the more challenging case of $\phi^4$ theory in 2D. For $m^2<0$ the theory undergoes a second order phase transition at some non-trivial coupling $\lambda_c(m^2)$ from an ordered phase $0 \leq \lambda < \lambda_c(m^2)$ to an unordered phase $\lambda > \lambda_c(m^2)$. The goal of this section is to study this phase transition for $m^2=-0.5$.\\
The bare 2-point correlation function in momentum space $G^0(\mathbf{p})$ has poles at $\sum_{i=1}^D \sin(\frac{p_i}{2})^2 = \frac{|m^2|}{4}$. In order to avoid these poles it is convenient to solve the first DSE for the 2-point correlation function, Eq.~(\ref{truncatedPhi4DSE}), in the form $G^{-1}=G_{0}^{-1} - \Sigma$ leading to the root finding problem:
\begin{equation}
\begin{split}
\Gamma^{(2)} & ( \mathbf{p} ) - G_{0}^{-1}( \mathbf{p} ) + \frac{\lambda}{2} \int_k \frac{1}{ \Gamma^{(2)}( \mathbf{k} )} - \\
 & \frac{\lambda}{6} \int_{k,q} \frac{ \Gamma( \mathbf{p+k+q}, \mathbf{q}, \mathbf{k} ) }{\Gamma^{(2)}( \mathbf{k} ) \Gamma^{(2)}( \mathbf{q} ) \Gamma^{(2)}( \mathbf{p+k+q} )}  = 0,
\end{split}
\label{inverseDSEtwopoint_momentum}
\end{equation}
where the DSE for the vertex function $\Gamma$ is still given by the Fourier transform of the second equation in (\ref{truncatedPhi4DSE}).\\
In contrast to Sec.~\ref{sec:phi4HAM} it is impossible to apply the self-consistency of Fig.\;\ref{fig:selfconsistencyloop} straightforwardly as the high-dimensional object $\Gamma$ can no longer be stored and manipulated to high accuracy.
In order to solve the coupled equations the stochastic evaluation of the tree expansion is used in the self-consistency of Fig.\;\ref{fig:combinedsolver}.
How the algorithm discussed for the 1D example in Sec.~\ref{sec:phi4_Gamma} is used in the 2D case will be explained in the following.
\subsection{Diagrammatic Monte Carlo setup}
In the $n$-th self-consistency step, cf. Fig.\;\ref{fig:combinedsolver}, the solution of Eq.~(\ref{inverseDSEtwopoint_momentum}) is denoted as $\Gamma^{(2),(n+1/2)}$. This solution is found by using the HAM on (\ref{inverseDSEtwopoint_momentum}) for fixed $\Gamma$ which leads to the $m$-th order deformation equation (the loop index $n$ is omitted in $u_{\Gamma^{(2),(n+1/2)}, m}$):
\begin{equation}
\begin{split}
& u_{\Gamma^{(2)}, m} (\mathbf{p})  = \chi_m u_{\Gamma^{(2)}, m-1} (\mathbf{p}) - h \left[   u_{\Gamma^{(2)}, m-1} (\mathbf{p}) - \right. \\
& \left. \;\; \tilde{\chi}_m G_{0}^{-1}( \mathbf{p} ) + \frac{\lambda}{2} H[ u_{\Gamma^{(2)} }, m ] -  \frac{\lambda}{6} S[ u_{\Gamma^{(2)} }, m ]( \mathbf{p} )    \right] 
\end{split}
\label{diagMCSolverI_II}
\end{equation}
where $u_{\Gamma^{(2)}, m}( \mathbf{p} ) = \frac{1}{m!} \frac{\text{d}^m}{\text{d} \text{q}^m} \left. \phi( \mathbf{p}, q ) \right|_q=0$, and
\begin{equation}
\begin{split}
H & [ u_{\Gamma^{(2)} }, m ] = \frac{1}{(m-1)!} \int_k \left. \frac{ \text{d} ^{m-1}}{\text{d} \text{q} ^ {m-1}} \frac{1}{\phi(\mathbf{k},q)} \right|_{q=0} \, ,\\
S & [ u_{\Gamma^{(2)} }, m ]( \mathbf{p} ) =  \frac{1}{(m-1)!} \int_{k,l} \Gamma(\mathbf{p} + \mathbf{k} + \mathbf{l}, \mathbf{k}, \mathbf{l}) \times \\
&  \left. \frac{ \text{d} ^{m-1} }{ \text{d} \text{q} ^ {m-1}} \frac{1}{ \phi(\mathbf{k},q) \phi(\mathbf{l},q) \phi(\mathbf{p} + \mathbf{k} + \mathbf{l}, q) } \right|_{q=0} = \\
 & =  \int_{k,l} W[ u_{\Gamma^{(2)} }, m ] \, ( \mathbf{p}, \mathbf{k}, \mathbf{l} ) \Gamma(\mathbf{p} + \mathbf{k} + \mathbf{l}, \mathbf{k}, \mathbf{l}) \, .
\end{split}
\end{equation}
In the function $S[ u_{\Gamma^{(2)} }, m ]( \mathbf{p} )$ the vertex function $\Gamma$ is given by applying the HAM on the second DSE in (\ref{truncatedPhi4DSE}). 
According to the self-consistency loop in Fig.\;\ref{fig:combinedsolver} the tree expansion is constructed for $G^{-1}=\Gamma^{(2),(n-1/2)}$ in order to calculate $S[ u_{\Gamma^{(2)} }, m ]( \mathbf{p} )$ in the $n$-th self-consistency step.\\
In practice the function $S( \mathbf{p} )$ is calculated by discretizing the external momentum $\mathbf{p}$ on a grid and applying the diagMC algorithm of \ref{sec:phi4_Gamma} to sample all possible rooted tree diagrams with variable heights and now variable external momentum. The external momentum of the rooted tree diagrams is updated by importance sampling of the variables $\mathbf{p}$, $\mathbf{k}$ and $\mathbf{l}$ with respect to the integral weight $W[ u_{\Gamma^{(2)} }, m ] \, ( \mathbf{p}, \mathbf{k}, \mathbf{l} )$ and histograms are taken for the discrete external momentum points. These histograms are normalized and stored giving a discretized function $\hat{S}( \mathbf{p} )$ which is used to compute $\hat{u}_{\Gamma^{(2)}, m}( \mathbf{p} )$ on the same momentum grid. For calculating the $j$-th order deformation $u_{\Gamma^{(2)}, j}$ the deformations $u_{\Gamma^{(2)}, m}$ with $m<j$ are needed which can be retrieved by bilinear interpolation of the stored results $\hat{u}_{\Gamma^{(2)}, m}$. $H$ is calculated by an independent deterministic numerical integration algorithm which can be considered exact as the numerical errors are subleading to the stochastic diagMC errors.
\begin{figure}
\includegraphics[width=0.45\textwidth]{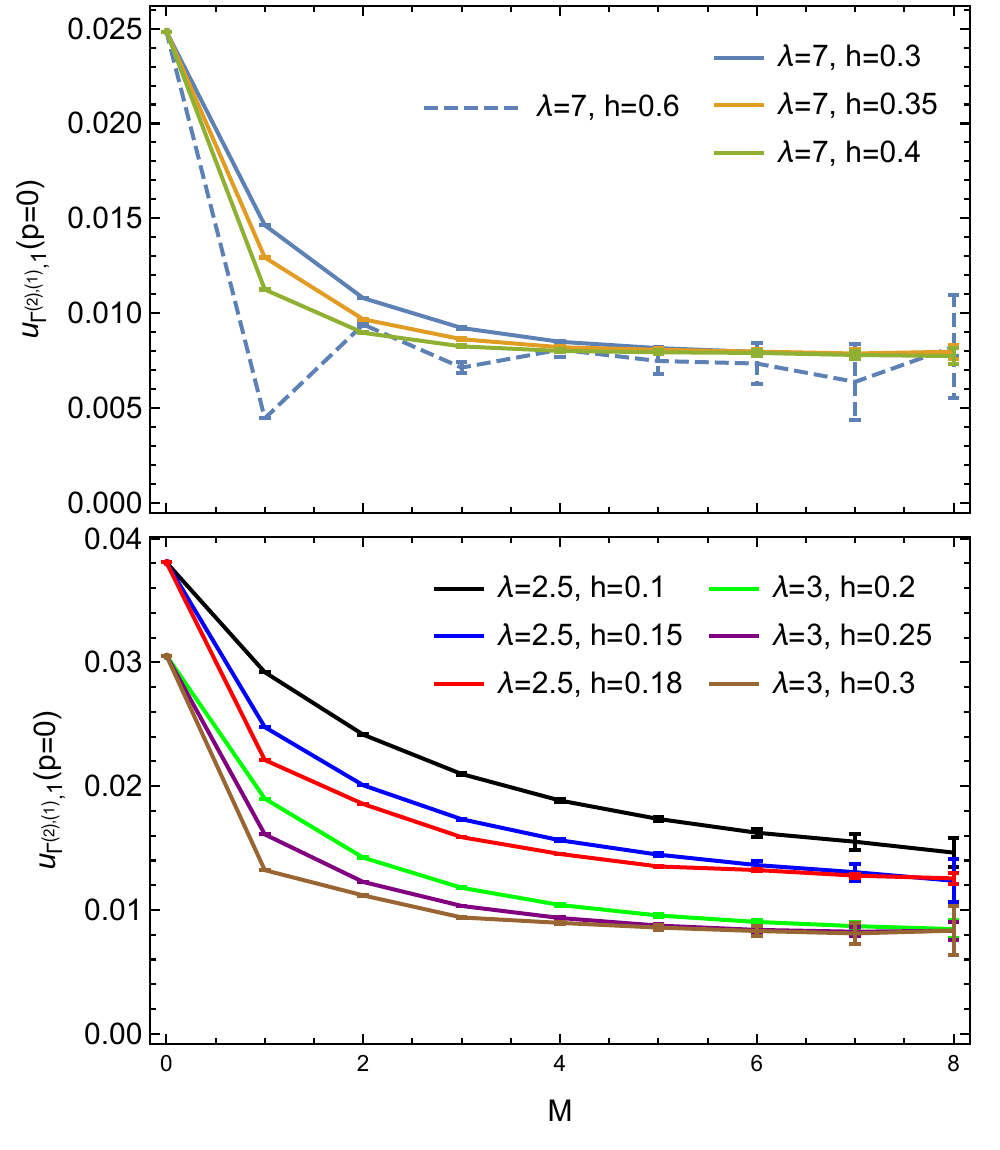}
\caption{The first deformation of $\Gamma^{(2)}$ in the first self-consistency loop, cf. Fig.\;\ref{fig:combinedsolver}, at zero external momentum for different truncation orders in the tree expansion of $\sum_{m=0}^{M} u_{\Gamma,m}$. Choosing $h$ carefully makes it possible to let the expansion converge in few orders and preventing a systematic error from the expansion order cutoff. While for $\lambda=7$ the tree expansion can already be truncated at $M=4$ for $\lambda=2.5$ the tree expansion has to be considered up to order $M=8$.}
\label{fig:deformationzeromomentum}
\end{figure}
\subsection{Further approximations}
Before discussing results for the full solution of the coupled DSEs (\ref{truncatedPhi4DSE}) we introduce various truncations of $\Gamma$ which we will compare against in Sec~\ref{sec:five_results}. Setting $\Gamma=0$ in Eq.~(\ref{inverseDSEtwopoint_momentum}) transforms the non-linear integral equation into the non-linear algebraic equation
\begin{equation}
\begin{split}
m_R^2 - m^2 + \frac{\lambda}{2} \int_k \frac{1}{ \Gamma^{(2)}( \mathbf{k} ) } = 0 \\ \Gamma^{(2)}( \mathbf{k} ) = 4\sum_{i=1}^D \sin(\frac{k_i}{2})^2 + m_R^2  \, .
\end{split}
\end{equation}
This equation can be easily solved by tabulating the integral for different $m_R$ and using standard numerical methods for solving algebraic non-linear equations.\\
Another simple truncation is to take $\Gamma = \text{const.} \neq 0$ in momentum space. In order to find a non-trivial fixed point the dimensionless renormalized coupling constant in 2D $\tilde{\lambda}_R = \Gamma(0,0,0) m_R^{-2}$ has to approach a non-trivial value if the system is tuned close to the phase transition. The next to leading order truncation to (\ref{truncatedPhi4DSE}) satisfying this condition is
\begin{equation}
\begin{split}
\Gamma^{(2)} ( \mathbf{p} ) = & \; G_{0}^{-1}( \mathbf{p} ) - \frac{\lambda}{2} \int_k \frac{1}{ \Gamma^{(2)}( \mathbf{k} )} + \\
  + & \frac{\lambda}{6} \int_{k,q} \frac{ \lambda_R[\Gamma^{(2)}] }{ \Gamma^{(2)}( \mathbf{k} ) \Gamma^{(2)}( \mathbf{q} ) \Gamma^{(2)}( \mathbf{p+k+q} ) } \\
  \lambda_R[\Gamma^{(2)}] = & \frac{\lambda}{1+\frac{3}{2}\lambda \int_k \left( \frac{1}{ \Gamma^{(2)}( \mathbf{k} )} \right)^2}.
 \end{split}
 \end{equation}
The latter equation corresponds to a resummed ladder expansion at zero external momentum.
This set of equations can be solved either by approximating $\Gamma^{(2)}( \mathbf{p} ) = 4\sum_{i=1}^D \sin(\frac{p_i}{2})^2 + m_R^2$, yielding a non-linear algebraic set of equations for $m_R$ and $\lambda_R$, or without further approximations by using the self-consistency loop, Fig.\;\ref{fig:selfconsistencyloop}, with ``Solver I" just calculating $\lambda_R$ and ``Solver II" the HAM for $\Gamma^{(2)}( \mathbf{p} )$.\\
\subsection{Numerical sign problem}
In order to obtain controlled diagMC simulation results  to the coupled integral equations of the truncated tower of DSEs the following issues have to be taken into account. The best choice of the free simulation parameters is as follows: The starting value of $\Gamma^{(2)}$, $\Gamma^{(2),(0)}$ in the self-consistency loop, Fig.\ref{fig:combinedsolver}, is taken to be the inverse of the 2-point correlation function obtained by the classical worm algorithm. The initial guess for the HAM root finding (\ref{diagMCSolverI_II}) in the $n$-th self-consistency loop is taken to be $u_{\Gamma^{(2),(n+1/2)}, 0}=\Gamma^{(2),(n-1/2)}$ and the initial guess for the tree expansion of the HAM is $u_{\Gamma,0}=\lambda_R[\Gamma^{(2),(n-1/2)}]$ a constant in momentum space. To determine the maximal order of the tree expansion $\sum_{m=0}^M u_{\Gamma,m}$ we calculate $u_{\Gamma^{(2),(1)}, 1}(\mathbf{p}=0)$, cf. Eq.~(\ref{diagMCSolverI_II}), for different expansion orders $M$ and various convergence control parameters $h$, cf. (\ref{deformingGamma}). Examples of these calculations are shown in Fig.\,\ref{fig:deformationzeromomentum}. The convergence control parameter must be chosen such that the convergence is as fast as possible, i.e. the expansion order is as low as possible. For $\lambda=7$ the choice $h=0.4$ is optimal while a higher value $h=0.6$ leads to an oscillating solution with high error bars. If $h$ is set to too small values (like for the $\lambda=2.5$, $h=0.1$ case) systematic errors in the calculation of the deformations of $\Gamma^{(2)}$ are introduced through the omission of higher order deformations.\\
\begin{figure}[t]
\includegraphics[width=0.45\textwidth]{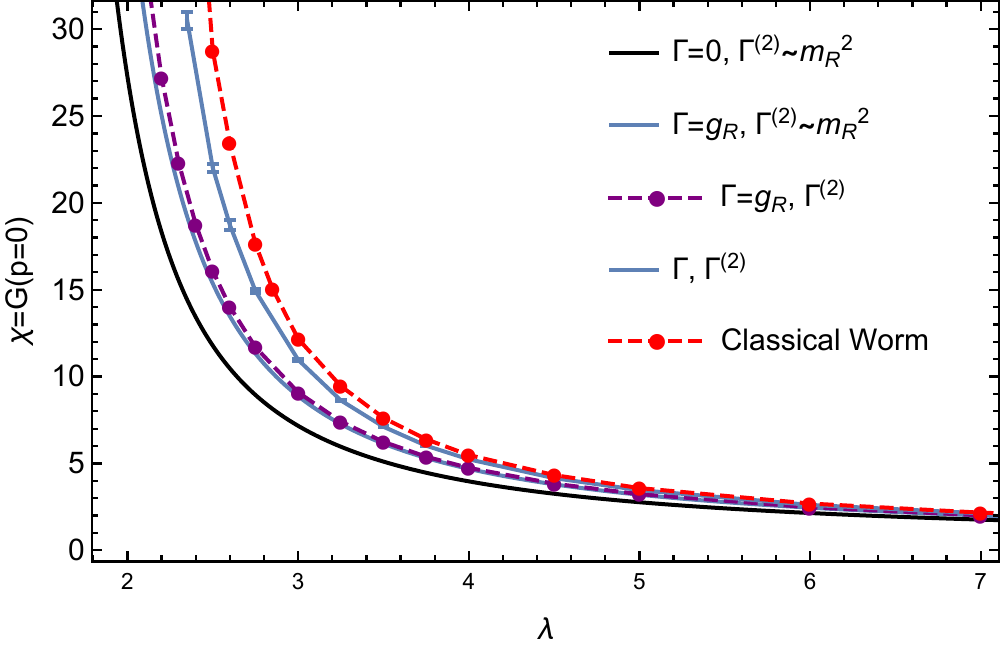}
\caption{The results from various approximations to the coupled DSEs (\ref{truncatedPhi4DSE}) compared to the classical worm algorithm.}
\label{fig:compare_susc}
\end{figure}
The average sign $\langle s \rangle$ of the diagMC integration also depends on the convergence control parameter. While for $\lambda=7$, $h=0.3$, $\langle s \rangle =  -0.18$, it goes down to $\langle s \rangle =  -0.09$ for $h=0.4$ and drops further to $-0.03$ for $h=0.6$ in the case of an oscillating convergence. As already noted in Sec.~\ref{sec:phi4_Gamma} the sign problem originates from the different signs in front of the linear and quadratic term with respect to $\Gamma$ in the DSE for the vertex function, cf. Eq~(\ref{truncatedPhi4DSE}), and therefore all rooted tree diagrams can be classified by having either a positive or negative contribution to $S( \mathbf{p} )=S^{+}( \mathbf{p} )-\left| S^{-}( \mathbf{p} ) \right|$ in Eq.~(\ref{diagMCSolverI_II}). Note that $\Gamma^{(2)}$ in the tree expansion is positive definite. The number of rooted tree diagrams in order $m$ grows exponentially with $m$. Therefore, without the weights of the rooted tree diagrams cancelling this exponential growth,
\begin{equation}
\left| S^{\pm}_m( \mathbf{p} ) \right| = \int_{k,l} W[ u_{\Gamma^{(2)} }, m ] \, ( \mathbf{p}, \mathbf{k}, \mathbf{l} ) \left| u^{\pm}_{\Gamma, m} (\mathbf{p}, \mathbf{k}, \mathbf{l} ) \right|
\end{equation}
will grow exponentially. Here, $u^{\pm}_{\Gamma, m}$ denotes the sum of all rooted tree diagrams contributing with a positive (negative) sign in the tree expansion of $u_{\Gamma, m}$.
Assuming convergence in the form $S_{m}( \mathbf{p} ) \rightarrow 0$ for $m \rightarrow \infty$ while $S_{m}( \mathbf{p} )$ having a definite sign for all $m$, which is actually the case for a carefully chosen $h$ as can be deduced from Fig.\;\ref{fig:deformationzeromomentum}, both big numbers $S^{\pm}_m( \mathbf{p} )$ have to be computed to very high precision in order for their difference to have  a high precision. This also explains the fact why $\langle s \rangle$ is smaller for $h=0.4$ compared to $h=0.35$ as convergence sets in faster for $h=0.4$. Thus, with carefully chosen simulation parameters it is possible to extract meaningful results before the statistical errors become too large due to the numerical sign problem. If one has oscillating convergence the sign problem makes it, as expected, impossible to extract controlled results in a realistic simulation time.\\
Apart from the cancellation of terms in the calculation of $S( \mathbf{p} )$ more cancellations occur in the calculation of the deformations $u_{\Gamma^{(2)},m}$. Assuming convergence of the self-consistency loop, i.e. $\Gamma^{(2)(n+1/2)} \approx \Gamma^{(2)(n-1/2)}$, it follows that $u_{\Gamma^{(2)(n+1/2)}, 1} \approx 0$. But $u_{\Gamma^{(2)(n+1/2)}, 1} \approx 0$ is obtained by an almost perfect cancellation of the individual contributions in (\ref{diagMCSolverI_II}) and clearly the absolute error $\Delta u_{\Gamma^{(2)(n+1)}, 1}$ is given by the statistical error of the diagMC calculation of $S$. As this statistical error is almost constant with respect to the self-consistency loop index $n$ the relative error $\delta u_{\Gamma^{(2)(n+1/2)}, 1} = \frac{\Delta u_{\Gamma^{(2)(n+1/2)}, 1}}{u_{\Gamma^{(2)(n+1/2)}, 1}}$ diverges. It is hence the sign problem that poses the strongest limitation on approaching the critical point more closely.
\subsection{Results}
\label{sec:five_results}
In order to compare the quality of the truncations of the infinite tower of DSEs the susceptibility $\chi=\frac{1}{\Gamma^{(2)}(0)}=G(0)$ is calculated in each setup and compared to the results from the classical worm algorithm \cite{ClassicalWorm}. As the worm algorithm works only on a finite lattice one must ensure that the system size is much larger than the correlation length. The results are shown in Fig.\,\ref{fig:compare_susc}. The quality of the truncation increases smoothly with the number of terms taken exactly into account. Deviations with the classical worm remain nevertheless visible, which is attributed to the the omission of higher order vertices such as $\Gamma^{(6)}$, but we checked that the data show consistency (albeit within the large error bars) with the critical exponent $\gamma=7/4$ for the susceptibility.
%
%
\section{Conclusion and Outlook}
\label{sec:conclusion}
In this paper we have treated the Dyson-Schwinger equations (DSEs) as integral equations posing a root finding problem, which we solved by introducing a rooted tree expansion in the Homotopy Analysis Method (HAM) framework, which has better convergence properties than fixed point iterations. We introduced a Monte Carlo sampling procedure to deal with the proliferation of branches and leafs in the tree expansion. 
Storing high-dimensional objects such as 4-point correlation functions, which naturally appear in the DSEs, can be avoided as long as one is not directly interested in the full knowledge of these quantities. We compared this tree expansion with the skeleton series expansion of the Luttinger-Ward functional in bold diagrammatic Monte Carlo and showed that the convergence properties of this new expansion are superior for the case of $\phi^4$ theory. We could go up to correlation length $5$ for the 2D model where further increase is limited by the sign problem.\\
In future work, the following two major questions must be addressed: The first one is about the quality of the truncation of the DSEs, which has partly already been addressed in previous works on DSEs \cite{DSEQCD} and within the functional renormalization group community \cite{fRGDelamotte, fRGStatMech, fRGFermi}. It has been shown that truncating the average effective action in the fRG approach by using a low order derivative and field expansion yields already good results for the critical exponents of $\phi^4$ models \cite{fRGStatMech}. We expect this to hold for the DSEs as well since both methods have to yield exactly the same results in the absence of approximations \cite{EllwangerDSEvs.fRG}.\\
The second major question concerns the sampling and manipulation of higher order vertices in case one decides to keep them in the expansion. In principle the ideas developed in this paper can be generalized to construct a tree expansion of higher order vertices: There will be another tree each time a higher order vertex appears. If the sign problem remains manageable, the answer can in principle be obtained, but this seems questionable when fluctuations dominate. \\
Interestingly, the functional form of the integro-differential formulation of the DSEs (cf. \cite{PelsterGlaum}) is such that another approach is feasible: $n$-point vertices with $n>4$ can be written as functional derivative terms of the 4-point vertex with respect to the full 2-point correlation function. In the rooted tree expansion of the HAM for the 4-point vertex function the functional derivative terms can be included naturally, as is shown in Appendix \ref{Integro-differential}. Hence, high-dimensional objects do not appear and this may be an interesting avenue for future work.\\
Finally, we note that the DSE formulation can equally well be applied to fermionic systems. It may in particular be interesting to study the Hubbard model near half filling, where the breakdown of the Luttinger-Ward functional and the convergence of the bold diagrammatic Monte Carlo approach to an unphysical solution have been reported~\cite{Kozik}. Our approach should not suffer from these problems.
\section{Acknowledgement}
\label{sec:ack}
This work was supported by FP7/ERC Starting Grant No. 306897 (``QUSIMGAS''). We are grateful to P. Kroi{\ss} for
useful discussions. TP acknowledges the help on technical questions from P. Kroi{\ss}. LP acknowledges support and hospitality from the Erwin Schr{\"o}dinger Institute, Vienna.
%
%

\newpage
%
%
\appendix
%
%
\section{Perturbative expansion in $\lambda$ for the DSEs}
\label{kappa2lambdaexp}
In this section we derive a perturbative expansion in $\lambda$ for the infinite tower of DSEs. The DSEs for the first three non-zero cumulants are given by
\begin{equation}
\begin{split}
\kappa_2 + \frac{\lambda}{6}\kappa_4 + \frac{\lambda}{2}\kappa_2^2&=1 \\
\kappa_4 + \frac{\lambda}{6}\kappa_6 + 2\lambda\kappa_2\kappa_4+\lambda\kappa_2^3&=0 \\
\kappa_6 + \frac{\lambda}{6}\kappa_8 + 5 \lambda \kappa_4 ^ 2 + 3 \lambda \kappa_2 \kappa_6 + 10 \lambda \kappa_2^2 \kappa_4 &=0.
\end{split}
\label{tower}
\end{equation}
The general idea can easily be illustrated by first setting $\kappa_6 = 0$. In this case the solution for the second equation of (\ref{tower}) is
\begin{equation}
\kappa_4 = \frac{-\lambda \kappa_2^3}{1+2 \lambda \kappa_2}.
\label{solution2}
\end{equation}
This solution can also be found by recursively plugging the second equation of $(\ref{tower})$ into itself,
\begin{align}
\kappa_4 = & -\lambda \kappa_2^3 - 2 \lambda \kappa_2 \kappa_4 = -\lambda \kappa_2^3 - 2 \lambda \kappa_2 ( -\lambda \kappa_2^3 - 2 \lambda \kappa_2 \kappa_4 )  \nonumber \\
 = & \dots = -\lambda \kappa_2^3 (1-2 \lambda \kappa_2 +  (-2 \lambda \kappa_2 )^2 - \dots) \label{solutionexpand} \\ 
 = &  \frac{-\lambda \kappa_2^3}{1+2 \lambda \kappa_2} \nonumber .
\end{align}
With this result the perturbative solution of (\ref{tower}) up to order $\mathcal{O}(\lambda^2)$ is given by 
\begin{equation}
\kappa_2  + \frac{\lambda}{2}\kappa_2^2 - \frac{\lambda^2}{6} \kappa_2^3 =1.
\end{equation}
In order to find all terms in such a perturbative expansion one has to know at which order in $\lambda$ the cumulant $\kappa_{2n}$ starts to contribute. From (\ref{solution2}) it can be seen that $\kappa_4$ is at least of $\mathcal{O}(\lambda)$. The leading order term for $\kappa_6$ (cf. \ref{tower}) is of $\mathcal{O}(\kappa_4 \lambda) = \mathcal{O}(\lambda^2)$. The general structure of the DSEs is such that the leading contribution of $\kappa_{2n}$ is $\mathcal{O}(\kappa_{2(n-1)}\lambda) = \mathcal{O}(\lambda^{n-1})$.\\
As an example, the perturbative expansion up to $\mathcal{O}(\lambda^4)$ can be constructed by considering the leading order term of $\kappa_6$.
\begin{equation}
\kappa_6 = -  10 \lambda \kappa_2^2 \kappa_4 + \mathcal{O}(\lambda^3).
\end{equation}
Consequently, $\kappa_4$ is given by
\begin{align}
\kappa_4 = & \frac{10}{6} \lambda^2 \kappa_2^2 \kappa_4 - 2 \lambda \kappa_2 \kappa_4 - \lambda \kappa_2^3 + \mathcal{O}(\lambda^4) \nonumber \\
    =& \frac{10}{6} \lambda^2 \kappa_2^2 (  - \lambda \kappa_2^3 ) - 2 \lambda \kappa_2 \left( - 2 \lambda \kappa_2 ( - \lambda \kappa_2^3 ) - \lambda \kappa_2^3 \right) \nonumber \\ 
      & - \lambda \kappa_2^3 + \mathcal{O}(\lambda^4) \\
    =& -\frac{17}{3} \lambda^3 \kappa_2^5 + 2 \lambda^2 \kappa_2^4 - \lambda \kappa_2^3 + \mathcal{O}(\lambda^4).
\end{align}
Plugging this result into the first equation of (\ref{tower}) yields the perturbative expansion in $\lambda$ up to order $\mathcal{O}(\lambda^4)$,
\begin{equation}
\kappa_2 + \frac{\lambda}{2}\kappa_2^2 - \frac{\lambda^2}{6} \kappa_2^3 + \frac{ \lambda^3 }{3} \kappa_2^4 - \frac{17}{18} \lambda^4 \kappa_2^5 + \mathcal{O}(\lambda^5) =1.
\end{equation}
\\
%
%

\section{Integro-differential formulation of DSEs}
\label{Integro-differential}
In the functional integro-differential formulation of the infinite tower of DSEs \cite{PelsterGlaum} the equation for the 4-point vertex function can be written as
\begin{equation}
\begin{split}
\Gamma_{1,2,3,4} = & \; \text{r.h.s. Eq. (\ref{truncatedPhi4DSE}) } - \frac{\lambda}{3} \int_{5,6,7} A_{1,5,6,7 } \frac{\delta \Gamma_{5,2,3,4}}{\delta G_{6,7}} \nonumber \\
   & + \frac{\lambda}{6} \int_{5,\dots,11} B_{1,5,\dots,11} \times \\
   & \hspace{25mm} \Gamma_{6,7,10,11} \frac{\delta \Gamma_{5,2,3,4}}{\delta G_{8,9}}, \nonumber
\end{split}
\end{equation}
where the kernel functions are given by
\begin{equation}
\begin{split}
A_{1,5,6,7 } =& \; G_{1,5}G_{1,6}G_{1,7} \nonumber \\
B_{1,5,\dots,11} =& \; G_{1,5}G_{1,6}G_{1,7} G_{8,10} G_{9,11}.
\end{split}
\end{equation}
The Homotopy Analysis Method can be applied to this equation leading to the $m$-th order deformation equation 
\begin{equation}
\begin{split}
u_{m,1,2,3,4} = & \; \text{r.h.s Eq. (\ref{deformingGamma}) } +  \frac{h \lambda}{3} \int_{5,6,7} \!\!\! A_{1,5,6,7} \frac{\delta u_{m-1,5,2,3,4}}{\delta G_{6,7}} \\
 - \frac{h \lambda}{6} & \int_{5,\dots,11} \!\!\!\!\!\! B_{1,5,\dots,11} \sum_{k=0}^{m-1} u_{m-1-k,6,7,10,11} \frac{\delta u_{k,5,2,3,4}}{\delta G_{8,9}}. \nonumber
\end{split}
\end{equation}
This equation is again the starting point for the tree expansion where all references to $u_j$, $j<m$ will be eliminated in the definition of the $m$-th order deformation equation. To see how the functional derivatives are handled consider for example the following term in the tree expansion of $u_3$:  
\begin{equation}
\begin{split}
 - &\frac{h^3 \lambda^3}{36}  \int_{5,\dots,11} \!\!\!\!\!\! B_{1,5,\dots,11} u_{0;6,7,10,11} \frac{\delta}{\delta G_{8,9}} \left[ \int_{12,13,14} \!\!\!\!\!\! A_{5,12,13,14} \times \right. \\
 & \left. \frac{\delta }{\delta G_{13,14}} \left( \delta(12-2) \int_{15,16} G_{12,15} G_{12,16} u_{0;15,16,3,4} \right) \right]. \nonumber
\end{split}
\end{equation}
The functional derivatives can be evaluated easily by using $\frac{ \delta u_0 }{ \delta G }=0$ and $\frac{\delta G_{1,2}}{\delta G_{3,4}} = \frac{1}{2} ( \delta_{1,3} \delta_{2,4} + \delta_{1,4} \delta_{2,3} )$. Due to the product rule for functional derivatives three different terms are generated.
\begin{equation}
\begin{split}
-\frac{h^3\lambda^3}{18} & \int_{5,\dots,11} u_{0;8,9,3,4} u_{0;6,7,10,11} \times \\
& \left[ G_{1,5}G_{1,6}G_{1,7}G_{2,10}G_{9,11}G_{5,2}^2G_{5,8} \right. \\ 
& +\;2 G_{1,5}G_{1,6}G_{1,7}G_{5,10}G_{2,11}G_{5,2}G_{5,8}G_{5,9} \\
& +\left. G_{1,5}G_{1,6}G_{1,7}G_{5,10}G_{8,11}G_{5,2}^2 G_{2,9}\right] \nonumber
\end{split}
\end{equation}
These terms can be accounted for in the stochastic evaluation of the tree expansion by extending the configuration space with all possible rooted tree diagrams generated by the functional derivative terms.


\begin{thebibliography}{99}
\bibitem{RecentDevelopmentsPIMC} L. Pollet, Rep. Prog. Phys. \textbf{75}, 094501 (2012).
\bibitem{ComparisonMCExperiment}  S. Trotzky, L. Pollet, F. Gerbier, U. Schnorrberger, I. Bloch, N. V. Prokof'ev, B. V. Svistunov and M. Troyer, Nature Physics \textbf{6}, 998 (2010).
\bibitem{HiggsBoseHubbard} L. Pollet and N. V. Prokof'ev, Phys. Rev. Lett. \textbf{109}, 010401 (2012).
\bibitem{FateVacancySuperSolid} M. Boninsegni, A. B. Kuklov, L. Pollet, N. V. Prokof'ev, B. V. Svistunov and M Troyer, Phys. Rev. Lett. \textbf{97}, 080401 (2006).
\bibitem{SFGrainBoundaries} L. Pollet, M. Boninsegni, A. B. Kuklov, N. V. Prokof'ev, B. V. Svistunov and M. Troyer, Phys. Rev. Lett. \textbf{98}, 135301 (2007).
\bibitem{NuclearShell} C. W. Johnson, S. E. Koonin, G. H. Lang, and W. E. Ormand, Phys. Rev. Lett. \textbf{69}, 3157 (1992).
\bibitem{HybridMonteCarlo} S. Duane, A. D. Kennedy, B. J. Pendleton and D. Roweth, Phys. Lett. B \textbf{195}, 216 (1987).
\bibitem{UnitaryFermionsI} A. Bulgac, J. E. Drut and P. Magierski, Phys. Rev. Lett. \textbf{96}, 090404 (2006).
\bibitem{UnitaryFermionsII} E. Burovski, N. Prokof'ev, B. Svistunov and M. Troyer, Phys. Rev. Lett. \textbf{96}, 160402 (2006).
\bibitem{detQMCLatticeFermions}  R. Blankenbecler, D. J. Scalapino and R. L. Sugar, Phys. Rev. D \textbf{24}, 2278 (1981).
\bibitem{KaneMeleHubbardAssad} M. Hohenadler, T. C. Lang and F. F. Assaad, Phys. Rev. Lett. \textbf{106}, 100403 (2011).
\bibitem{KaneMeleHubbardWu} D. Zheng, G.-M. Zhang and C. Wu, Phys. Rev. B \textbf{84}, 205121 (2011).
\bibitem{GaugeFields} F.F. Assaad and T. Grover, Phys. Rev. X {\bf 6}, 041049 (2016).
\bibitem{ImpuritySolvers} E. Gull, A. J. Millis, A. I. Lichtenstein, A. N. Rubtsov, M. Troyer and P. Werner, Rev. Mod. Phys. \textbf{83}, 349 (2011).
\bibitem{DiffusionMonteCarloElectronGroundState} W. M. C. Foulkes, L. Mitas, R. J. Needs and G. Rajagopal, Rev. Mod. Phys. \textbf{73}, 33 (2001).
\bibitem{FICQMCcode} G. H. Booth, A. J. W. Thom and A. Alavi, J. Chem. Phys. \textbf{131}, 054106 (2009).
\bibitem{FICQMCapplication} G. H. Booth, A. Grüneis, G. Kresse and A. Alavi, Nature \textbf{493}, 365 (2013).
\bibitem{SignProblem} M. Troyer and U.-J. Wiese, Phys. Rev. Lett. \textbf{94}, 170201, (2005).
\bibitem{FermiPolaron} N. Prokof'ev and B. Svistunov, Phys. Rev. B \textbf{77}, 020408 (2008).
\bibitem{FermiPolaron3DMass} P. Kroiss and L. Pollet, Phys. Rev. B \textbf{91}, 144507 (2015).
\bibitem{FermiPolaron2DVlietinck} J. Vlietinck, J. Ryckebusch and K. Van Houcke, Phys. Rev. B \textbf{89}, 085119 (2014).
\bibitem{FrustratedSpins} S. A. Kulagin, N. Prokof'ev, O. A. Starykh, B. Svistunov, and C. N. Varney, Phys. Rev. Lett. \textbf{110}, 070601 (2013).
\bibitem{SpinIce} Y. Huang, K. Chen, Y. Deng, N. Prokof'ev and Boris Svistunov, Phys. Rev. Lett. \textbf{116}, 177203 (2016).
\bibitem{Hubbard} Y. Deng, E. Kozik, N. Prokof'ev and B. Svistunov, EPL \textbf{110}, 57001 (2015). 
\bibitem{AnisotropicHubbard} J. Gukelberger, E. Kozik, L. Pollet, N. Prokof'ev, M. Sigrist, B. Svistunov and M. Troyer, Phys. Rev. Lett. \textbf{113}, 195301 (2014).
\bibitem{PseudoGapHubbard} W. Wu, M. Ferrero, A. Georges and Evgeny Kozik, arXiv:1608.08402 (2016).
\bibitem{DiracLiquid} I. Tupitsyn and N. Prokof'ev, arXiv:1608.00133 (2016).
\bibitem{ColoumbElectronPhonon} I. S. Tupitsyn, A. S. Mishchenko, N. Nagaosa and N. Prokof'ev, Phys. Rev. B \textbf{94}, 155145 (2016).
\bibitem{DualFermionGull} S. Iskakov, A. E. Antipov and E. Gull, Phys. Rev. B \textbf{94}, 035102 (2016).
\bibitem{DualFermionGukelberger} J. Gukelberger, E. Kozik and H. Hafermann, arXiv:1611.07523 (2016).
\bibitem{ResonantFermiGas}  K. Van Houcke, F. Werner, E. Kozik, N. Prokof'ev, B. Svistunov, M. J. H. Ku, A. T. Sommer,	L. W. Cheuk, A. Schirotzek and M. W. Zwierlein, Nature Physics \textbf{8}, 366 (2012).
\bibitem{FermiPolaron2DKroiss} P. Kroiss and L. Pollet, Phys. Rev. B \textbf{90}, 104510 (2014).
\bibitem{GrassmanizationIsing} L. Pollet, M. N. Kiselev, N. V. Prokof'ev and B. V. Svistunov, New J Phys {\bf 18}, 113025 (2016).
\bibitem{Phi4Buividovich} P. V. Buividovich, Nuclear Physics B \textbf{853}, 688-709 (2011).
\bibitem{BoldDiagMonteCarlo} N. Prokof'ev and B. Svistunov, Phys. Rev. Lett. \textbf{99}, 250201 (2007).
\bibitem{LuttingerWard} J. M. Luttinger and J. C. Ward,, Phys. Rev. \textbf{118}, 1417 (1960).
\bibitem{HAMbook} S. J. Liao, \textit{Beyond Perturbation: Introduction to the Homotopy Analysis Method} (Chapman and Hall/CRC, 2003).
\bibitem{HAMbook2} S. J. Liao, \textit{Homotopy Analysis Method in Nonlinear Differential Equations} (Berlin and Beijing, Springer and Higher Education Press, 2012).
\bibitem{HAMbook3} K. Vajravelu and R. A. Van Gorder, \textit{Nonlinear Flow Phenomena and Homotopy Analysis} (Heidelberg and Beijing, Springer and Higher Education Press, 2012).
\bibitem{BellPolynomialRef} J. Riordan, \textit{Introduction to Combinatorial Analysis} (Princeton University Press, 2014).
\bibitem{PelsterGlaum} A. Pelster and K. Glaum, Physica A \textbf{335}, 455-486 (2004).
\bibitem{DSEQCD} R. Alkofera and L. von Smekalb, Physics Reports \textbf{353}, 281-465 (2001).
\bibitem{fRGDelamotte} B. Delamotte in {\it Renormalization Group and Effective Field Theory Approaches to Many-Body Systems},  Eds. A. Schwenk and J. Polonyi, (Springer Berlin Heidelberg, ISBN 978-3-642-27320-9, 2012), arXiv:cond-mat/0702365 (2007)
\bibitem{fRGStatMech} J. Berges, N. Tetradis and C. Wetterich, Phys. Rept. \textbf{363}, 223-386 (2002).
\bibitem{fRGFermi} W. Metzner, M. Salmhofer, C. Honerkamp, V. Meden and Kurt Sch{\"o}nhammer, Rev. Mod. Phys. \textbf{84}, 299 (2012).
\bibitem{EllwangerDSEvs.fRG} U. Ellwanger, M. Hirsch, A. Weber, Eur. Phys. J. C \textbf{1}, 563-578 (1998).
\bibitem{ClassicalWorm} N. Prokof'ev and Boris Svistunov, Phys. Rev. Lett. \textbf{87}, 160601 (2001).
\bibitem{Kozik} E. Kozik, M. Ferrero, and A. Georges, Phys. Rev. Lett. {\bf 114}, 156402 (2015).
\end{thebibliography}
\end{document}